\documentclass[11pt]{article}
\usepackage[]{natbib}
\usepackage[margin=1in]{geometry}
\usepackage{amsmath,amssymb,amsthm,mathtools,bm}
\usepackage{mathabx,mathrsfs}
\usepackage{amsmath,amssymb,amsfonts}
\usepackage{mathtools}
\usepackage{bm}
\usepackage{algorithm}
\usepackage{algpseudocode}
\usepackage{microtype}
\usepackage{placeins}
\usepackage{enumitem}
\usepackage{hyperref}
\usepackage{booktabs}
\usepackage{setspace}
\usepackage{comment}
\newtheorem{theorem}{Theorem}
\newtheorem{definition}{Definition}
\newtheorem{proposition}{Proposition}
\newtheorem{corollary}{Corollary}

\newtheorem{assumption}{Assumption}
\newtheorem{remark}{Remark}
\newcommand{\blind}{1}
\usepackage{xcolor}

\newcommand{\PP}{\mathbb{P}}
\newcommand{\1}{\mathbf{1}}
\newcommand{\diag}{\operatorname{diag}}
\newcommand{\argmin}{\operatorname*{arg\,min}}

\newcommand{\mix}{\operatorname{mix}}

\newcommand{\base}{\operatorname{base}}

\def\supl{^{\scriptscriptstyle (\ell)}}
\def\sublk{_{\scriptscriptstyle \ell,k}}
\def\sublkp{_{\scriptscriptstyle \ell,k'}}
\def\supm{^{\scriptscriptstyle [m]}}
\def\suplm{^{\scriptscriptstyle (\ell)[m]}}
\def\supzerom{^{\scriptscriptstyle (0)[m]}}

\newcommand{\Dsc}{\mathcal{D}}

\def\trans{^{\scriptscriptstyle \sf T}}
\newcommand{\GUARD}{\mathrm{GUARD}}

\newcommand{\bbeta}{\beta}

\newcommand{\bB}{B}

\newcommand{\bA}{A}
\newcommand{\bX}{X}
\newcommand{\Sig}{\Sigma_0}

\newcommand{\Snorm}[1]{\lVert #1 \rVert^2_{\Sig}}

\usepackage{enumitem}

\def\subfut{_{\scriptscriptstyle \sf fut}}
\def\bbetafut{\bbeta\subfut}
\def\Bscrfut{\mathscr{B}\subfut}
\def\Rworst{\mathcal{R}_{\scriptscriptstyle \sf worst}}
\def\Ept{\Ebb_{\Pt}}

\def\El{\Ebb_{\ell}}
\def\Et{\Ebb_0}
\def\Ebb{\mathbb{E}}
\def\Pbb{\mathbb{P}}
\def\Qbb{\mathbb{Q}}

\def\Rbb{\mathbb{R}}
\def\supzero{^{\scriptscriptstyle (0)}}
\def\Pnew{\Pbb_{\rm\scriptscriptstyle new}}
\def\Pt{\Pbb\supzero}
\def\Pl{\Pbb\supl}
\def\betanew{\beta_{\rm \scriptscriptstyle new}}

\def\PbbhatLl{\widehat\Pbb_{\scriptscriptstyle \Lsc}\supl}
\def\PbbhatUl{\widehat\Pbb_{\scriptscriptstyle \Usc}\supl}
\def\PbbhatLt{\widehat\Pbb_{\scriptscriptstyle \Lsc}\supzero}
\def\PbbhatUt{\widehat\Pbb_{\scriptscriptstyle \Usc}\supzero}

\def\PbbhatLlm{\widehat\Pbb_{\scriptscriptstyle \Lsc}\suplm}
\def\PbbhatUlm{\widehat\Pbb_{\scriptscriptstyle \Usc}\suplm}
\def\PbbhatLtm{\widehat\Pbb_{\scriptscriptstyle \Lsc}\supzerom}
\def\PbbhatUtm{\widehat\Pbb_{\scriptscriptstyle \Usc}\supzerom}

\def\PbbhatLkl{\widehat\Pbb_{\scriptscriptstyle \Lsc,k}\supl}
\def\PbbhatUkl{\widehat\Pbb_{\scriptscriptstyle \Usc,k}\supl}
\def\PbbhatLkt{\widehat\Pbb_{\scriptscriptstyle \Lsc,k}\supzero}
\def\PbbhatUkt{\widehat\Pbb_{\scriptscriptstyle \Usc,k}\supzero}

\def\PbbhatLklm{\widehat\Pbb_{\scriptscriptstyle \Lsc,k}\suplm}
\def\PbbhatUklm{\widehat\Pbb_{\scriptscriptstyle \Usc,k}\suplm}

\def\PbbhatUktm{\widehat\Pbb_{\scriptscriptstyle \Usc,k}\supzerom}

\def\Lsc{\mathcal{L}}
\def\Usc{\mathcal{U}}
\def\Isc{\mathcal{I}}
\def\Dsc{\mathcal{D}}

\def\Ll{\Lsc\supl}
\def\Ul{\Usc\supl}
\def\Dl{\Dsc\supl}

\def\subGUARD{_{\rm \scriptscriptstyle GUARD}}

\newcommand{\tot}{\operatorname{tot}}

\def\Gammahat{\widehat{\Gamma}}
\def\Sigmahat{\widehat{\Sigma}}
\newcommand{\Vhat}{\widehat V}

\newcommand{\Bhat}{\widehat B}
\newcommand{\betahat}{\widehat\beta}
\newcommand{\gammahat}{\widehat\gamma}
\def\bbeta{\boldsymbol{\beta}}

\def\trans{^{\scriptscriptstyle \sf T}}

\title{GUARD: Guided Uncertainty-Aware Robust Domain Transfer}
\date{}
\author{}

\usepackage{xr-hyper}
\begin{document}

\if1\blind
{
  \title{\bf Guided Uncertainty-Aware Robust Domain Transfer}
 \author{Xin Xiong$^{1}$, 
  Zijian Guo$^{2}$,
  Tianxi Cai$^{1,3}$
\bigskip \\ 
$^1${Harvard T.H. Chan School of Public Health, Boston, MA, USA} \\
$^2${Center for Data Science, Zhejiang University, Zhejiang, China} \\
$^3${Harvard Medical School, Boston, MA, USA} \\
\date{}
}
  \maketitle
} \fi

\if0\blind
{
  \bigskip
  \bigskip
  \bigskip
  \begin{center}
    {\LARGE\bf Guided Adversarial Robust Transfer Learning with Source Mixing}
\end{center}
  \medskip
} \fi

\bigskip
\begin{abstract}
Artificial intelligence models deployed as clinical decision support tools often suffer substantial performance degradation over time and across institutions due to covariate shift, concept drift, and cross‑system heterogeneity. Fully retraining complex models is frequently infeasible, particularly in EHR settings where labeled outcome data are scarce and regulatory constraints limit model modification. We propose GUARD (Guided and Uncertainty‑Aware Robustness to Domain shift), a unified statistical framework for updating an existing deployed model using limited labeled target data and multiple heterogeneous source populations while guarding against future distributional shifts. GUARD formulates robust multi‑source transfer learning as a semi‑supervised, adversarial optimization problem anchored at the current model, and employs double machine learning with cross‑fitting to obtain debiased, efficient estimates that leverage abundant unlabeled target covariates. It then constructs an uncertainty‑aware, guided group DRO estimator that combines source and target information while accounting for sampling variability in the source models. Our framework jointly provides principled robustness to domain shift, efficient semi‑supervised estimation, and valid statistical inference for multi‑source transfer of clinical prediction models. We demonstrate the utility of GUARD via extensive simulation experiments and a real world application to predicting future disease activity in rheumatoid arthritis from EHR data spanning more than a decade. GUARD recalibrates a pretrained model with only a small number of labeled records per year and remains accurate over multi-year horizons where target-only and standard transfer estimators degrade sharply.

\end{abstract}

\newpage
\section{Introduction}
The integration of artificial intelligence (AI) into clinical care has reached a pivotal juncture: AI-driven decision support tools are increasingly approved by the U.S. Food and Drug Administration (FDA) as Software as a Medical Device  \citep{muehlematter2023fda}. Commonly trained on electronic health record (EHR) data, these systems support tasks including risk stratification, early diagnosis, and personalized treatment. Yet ensuring their sustained effectiveness after deployment remains challenging \citep{abulibdeh2025deployment}. Most clinical AI models experience substantial effectiveness reductions within years of initial deployment \citep{vela2022temporal}, an ``aging'' phenomenon that can jeopardize patient safety and erode clinical trust. Naive solutions such as frequent retraining are often infeasible due to computational costs. In addition, AI algorithms approved as medical devices operate under regulatory frameworks that constrain how and when models may be retrained or modified, making it unclear how to properly maintain, update, and consistently evaluate and report their post-marketing performance \citep{muralidharan2024gaps}.

Performance degradation in EHR-based algorithms primarily stems from two often intertwined forms of drift: covariate shift, when feature distributions change due to evolving populations or varying health systems, and concept drift, when feature-outcome relationship shifts due to shifts such as updated clinical guidelines. Changes in coding systems, clinical practice, or disruptions such as the COVID pandemic can simultaneously cause both types of shift, compounding the challenge of long-term model reliability \citep{vela2022temporal}. Beyond such temporal instability, cross-institutional heterogeneity driven by differing coding standards,  practices, and patient demographics, introduces a further dimension of distributional shift that limits model transportability across environments.

These challenges are especially acute in EHR-based clinical settings, where labeled outcome data are costly to curate \citep{muralidharan2024gaps}. The scarcity of labeled data makes full retraining even less feasible, especially for complex models with a large number of parameters. A potentially more effective approach to updating AI models is through stable post-processing, where an existing model is refined using a limited amount of newly acquired labeled target-domain data alongside data from multiple related but heterogeneous source populations, representing, for example, historical time epochs from the same institution or patient cohorts from other health systems. During the updating process, it is also critical to adopt a drift-aware principle to ensure that the updated model is more robust to future distribution shifts. 

Transfer learning provides a natural paradigm for leveraging heterogeneous source populations to improve prediction on a data-scarce target domain \citep{laparra2021review}. Recent statistical contributions in transfer learning \citep[e.g.]{li2022translasso,tian2023transglm,li2023b} have rigorously formulated this idea. The existing literature typically requires identifying transferable sources, those with model parameters sufficiently similar to the target, while excluding others to prevent negative transfer. In practice, however, this notion of proximity  may be overly restrictive. Individual sources may be highly heterogeneous and each differs substantially from the target, yet collectively there may exist a mixture of the sources that approximates the target well. Such a source-mixture proximity perspective allows the target to borrow information from multiple sources without requiring each individual source to be close to the target. In addition, the standard transfer learning methods do not offer robustness against future shifts in the target distribution, leaving the re-trained model vulnerable to continued drift beyond the current target snapshot.

Group distributional robust optimization (DRO) offers an alternative framework motivated by this same robustness concern \citep{meinshausen2015maximin,sagawa2019distributionally, guo2023maximin}. For classification, \citet{guo2025statistical} formulate the corresponding problem under cross-entropy loss and develop optimization, statistical convergence, and inference theory. \citet{kim2026distributionally} extend conditional-mixture DRO to allow both adversarial source mixtures and Wasserstein perturbations of scarce target covariates, although their theory centers on a surrogate formulation rather than statistical guarantees or inference. Unlike transfer learning, which conditions on a fixed and observed target population, group DRO was designed for settings where no labeled data from the future deployement environment is available, and robustness is sought preemptively over the class of plausible future target populations formed from mixtures of the observed sources. Rather than restricting attention to transferable sources, DRO minimizes the worst-case expected loss over a pre-specified uncertainty set, often defined through convex combinations of source conditional distributions. While these methods provide valuable worst-case guarantees, they can be overly conservative and sacrifice accuracy on a specific future target. 

To resolve the tension between target-specific accuracy and distributional robustness, \citet{xiong2023gart} introduced Guided Adversarial Robust Transfer (GART), which uses a small amount of labeled target data to construct a smaller transferable uncertainty class for robust prediction on the observed target and nearby populations. Related approaches have further explored domain adaptation optimized for robustness in mixture populations \citep{zhan2024domain, mo2024minimax}. \citet{mansour2021theory} construct predictors under different mixtures of the source distributions and use limited target labels to select the one that performs best on the observed target. \citet{wegel2026hedging} frame multi-source transfer through weak monotonicity, using source risks to restrict target adaptation to selecting or hedging among predictors on the source Pareto frontier. These approaches use target information to guide selection or restrict the robust class, but they do not address semi-supervised target-guided transfer, where abundant unlabeled target covariates are available. They also typically treat source risks or source predictors as fixed, rather than accounting for heterogeneous source-estimation uncertainty.

Recent work has begun to study semi-supervised variants of related robust or multi-objective problems. \citet{awasthi2024semi} study semi-supervised group DRO for sparse-group settings, using unlabeled data and pseudo-labeling from informative dense groups to improve min--max performance across observed groups. Their framework relies on a pairwise similarity condition: for each sparse group, some dense group must have an optimal or near-optimal predictor that performs well on that sparse group. Relatedly, \citet{wegel2025sample} study semi-supervised multi-objective learning, showing that for Bregman losses, unlabeled data can reduce labeled sample complexity through pseudo-labeling. However, their objective is to learn Pareto-optimal trade-offs across observed tasks, rather than to guide transfer to a current or future target population. Thus, existing semi-supervised multi-source robust learning methods do not address the setting where both $\Pbb_X$ and $\Pbb_{Y|X}$ may differ across sources and the target, the working model may be misspecified, and and transferability may arise only through mixtures of heterogeneous sources. 

In this paper, we introduce GUARD (Guided and Uncertainty-Aware Robustness to Domain shift), a unified statistical framework for robust multi-source transfer learning with valid inference. GUARD is formulated as a semi-supervised, multi-source adversarial robust optimization problem that takes the current deployed model $f_0(X)$ as its initialization with local update as $\beta_0 f_0(X) + \bbeta_1\trans X$. It leverages both a small labeled sample and a large pool of unlabeled covariates from target domain as well as partially labeled data from $L$ source populations. For estimation, GUARD uses double machine learning (DML) with cross-fitting to obtain debiased and efficient estimators of target-projected source coefficients, allowing flexible nuisance function estimators and leveraging unlabeled target data. It then constructs an uncertainty-aware robust combination of source and target models via a guided group DRO criterion, which explicitly incorporates source-specific estimation uncertainty. For inference, we introduce a resampling procedure to account for variations in both the estimated source--target coefficients library and the learned robust weights. Together, these contributions make GUARD, to our knowledge, the first framework to provide simultaneously: 1) robust transfer from multiple heterogeneous sources anchored to the current deployed model; 2) efficient semi-supervised estimation leveraging unlabeled target covariates; 3) uncertainty-aware source aggregation that accounts for heterogeneous source precision; and 4) valid statistical inference for robust transfer learning estimators. 
The rest of the paper is organized as follows. Section~\ref{sec:guard_population} introduces the multi-source semi-supervised transfer setting and defines the target-projected source parameters. Section~\ref{sec:estimation} presents the GUARD estimation procedure and resampling-based inference procedure for GUARD. Section~\ref{sec:theory} establishes transferability, estimation, and inference guarantees. Section~\ref{sec:sim} evaluates GUARD in simulations under future generalization and current-target transfer settings. Section~\ref{sec:real-design0} applies GUARD to temporally robust disease-activity prediction in rheumatoid arthritis using longitudinal EHR data. We conclude with discussion in Section~\ref{sec:discussion}.

\section{GUARD Population Framework}
\label{sec:guard_population}

We first define the target-projected source parameters and the population GUARD objective.

\subsection{Notations and data setting}\label{subsec:notation_data_setting}

Let $[L]=\{1,\ldots,L\}$ and $[L]_+=\{0\}\cup[L]$, where
$\ell=0$ denotes the target population and $\ell\in[L]$ denotes the
sources. 
For each $\ell\in[L]_+$, let $\Pl_X$ denote the covariate distribution, $\Pl_{Y\mid X}$  the conditional distribution of $Y \mid X$, and $\Pl$ the joint law for source $\ell$. We write  
\begin{equation}
Y\supl = m_{\ell}(X\supl) + \epsilon\supl , \qquad \mbox{where }
\Ebb(\epsilon\supl \mid X\supl) = 0, \quad \mbox{and}\quad
m_\ell(x):=\Ebb_{\Pl}(Y\mid X=x),
\label{model-source}
\end{equation}
with some smooth unspecified $m_\ell(\cdot)$. Let $\El = \Ebb_{\Pl}$ for notational ease.

Let $f_0(X)$ denote the prediction rule currently available for deployment.
Rather than retraining a new rule from scratch, we consider a local update
based on an augmented feature map $A=A(X)\in\mathbb R^p$. The vector
$A(X)$ may include $f_0(X)$, selected components of $X$, and engineered or
learned transformations of $X$, including low-dimensional representations
from flexible models such as transformers. The update is
linear in this constructed feature space, with target coefficient 
\begin{equation}
    \beta_0 = \operatorname*{arg\,min}_{b\in\mathbb R^p}
\Ept
\bigl\{(Y-A\trans b)^2\bigr\} = \Sigma_0^{-1}\Ept(AY), \quad 
\mbox{where $\Sigma_0:= \Ept(AA\trans)$.}
\end{equation}
Although $m_0(X)$ is optimal under squared loss, we use a lower-dimensional linear update to avoid unstable flexible retraining with limited target labels while still recalibrating $f_0(X)$ and correct systematic drift. 
For each source $\ell$, we project its distribution on the same target update space: 
\begin{equation}
\beta_\ell = \operatorname*{arg\,min}_{b\in\mathbb R^p}
\Ebb_{X\sim \Pt_X,\,Y\mid X\sim \Pl_{Y\mid X}}
\bigl\{(Y-A \trans b)^2\bigr\} =
\Sigma_0^{-1}\Ebb_{X\sim \Pt_X}\{A m_\ell(X)\}.
\label{eq: source beta l}
\end{equation}
Thus, $\beta_l$ places the $\ell$th source outcome mechanism on the same target update scale as $\beta_0$, even when both $\Pbb_X$ and $\Pbb_{Y|X}$ differ across populations. We collect these target-projected coefficients in
\[
B=(\beta_0,\beta_1,\ldots,\beta_L)\in\mathbb R^{p\times(L+1)}.
\]

\subsection{Target-anchored source-mixture robustness}
\label{subsec:target_anchored_mixture}
GUARD represents a candidate future target population through a convex mixture of the observed target and source conditional outcome distributions, while keeping the current target covariate distribution fixed. Let
$\Delta_{L+1}=\{\gamma\in\mathbb R^{L+1}:\gamma_\ell\geq 0,\ \sum_{\ell=0}^L\gamma_\ell=1\}$ denote the probability simplex. For $\gamma\in\Delta_{L+1}$, define
\begin{equation}
\Pnew(\gamma) =\left( \Pt_X, \sum_{\ell=0}^L \gamma_\ell \Pl_{Y\mid X} \right).
\label{eq:pnew_gamma}
\end{equation}
Under squared loss, the optimal linear update under $\Pnew(\gamma)$ is
\begin{equation}
\betanew(\gamma)
=
\operatorname*{arg\,min}_{b\in\mathbb R^p}
\Ebb_{\Pnew(\gamma)}
\bigl\{(Y-A\trans b)^2\bigr\}
=
B\gamma .
\label{eq:beta_new_gamma}
\end{equation}

For post-deployment updating, the natural benchmark is the current target only update $\beta_0$. Define the target-guided excess-loss contrast
$$ h_\beta(X,Y) := (Y-A\trans\beta)^2-(Y-A\trans\beta_0)^2 .$$
A direct target-guided group DRO criterion
minimizes the worst-case expected excess loss
$\beta_{\mathrm{TG\mbox{-}DRO}}^\star
\in
\operatorname*{arg\,min}_{\beta\in\mathbb R^p}
\sup_{\gamma\in\Delta_{L+1}}
\Ebb_{\Pnew(\gamma)}\{h_\beta(X,Y)\}$.
However, this objective is self-anchoring with
$\beta_{\mathrm{TG\mbox{-}DRO}}^\star\equiv\beta_0$, since the target distribution belongs to $\{\Pnew(\gamma), \gamma \in \Delta_{L+1}\}$. 
A second limitation is that standard group DRO treats the target and all source directions as equally reliable. In practice, the projected coefficients $\beta_l$ are estimated with different precision because sources can differ in labeled sample sizes, outcome noise, covariate overlap with the target, and nuisance-estimation difficulty.
GUARD incorporates this heterogeneity in estimation uncertainty directly into the population target. Let
$\widehat B=(\widehat\beta_0,\widehat\beta_1,\ldots,\widehat\beta_L)$, where
$\widehat\beta_\ell$ is a consistent regular estimator of $\beta_\ell$.
We define the uncertainty matrix of the coefficient library as
\[
V=\Ebb\{(\widehat B-B)\trans\Sigma_0(\widehat B-B)\},\qquad \mbox{and}\qquad
\gamma\trans V\gamma
=\Ebb\{\|\widehat B\gamma-B\gamma\|_{\Sigma_0}^2\}.
\]
Unlike the coefficient library $B$, $V$ reflects finite-sample estimation uncertainty and therefore depends on the source-specific sample sizes and noise levels. GUARD uses $\gamma\trans V\gamma$, which measures the target-scale estimation uncertainty of the mixture coefficient $B\gamma$, to favor mixtures that are both close to the current target update $\beta_0$ and reliably estimated. 

\subsection{GUARD as uncertainty-calibrated perturbed-control DRO}
\label{subsec:guard_dro_definition}

To account for source-estimation uncertainty, GUARD introduces local perturbations of each conditional outcome law along the working update directions. For each $\ell\in[L]_+$ and each perturbation vector $\varepsilon_\ell\in\mathbb R^p$, let $\Qbb_{\varepsilon_\ell,Y\mid X}^{(\ell)}$ denote the perturbed law satisfying
\begin{equation}
    \Ebb_{\Qbb_{\varepsilon_\ell}^{(\ell)}}[Y\mid X]
=
m_\ell(X)+A\trans\varepsilon_\ell .
\label{eq: Ql perturb def}
\end{equation}
This perturbation shifts the
target projected coefficient by $\varepsilon_\ell$:
\[
\beta_\ell^{Q}:=\arg\min_{b\in\mathbb R^p}
\Ebb_{\substack{X\sim \Pt_X, \,
Y\mid X\sim \Qbb_{\varepsilon_\ell,Y\mid X}^{(\ell)}}}
\left[(Y-A\trans b)^2 \right] =\beta_\ell+\varepsilon_\ell.
\]
Define the corresponding perturbed future population
\[\textstyle
\Pbb_{\gamma,E}=\left(\Pt_X, \sum_{\ell=0}^L
\gamma_\ell \Qbb_{\varepsilon_\ell,Y\mid X}^{(\ell)}\right), \quad
\mbox{for $E=(\varepsilon_0,\ldots,\varepsilon_L)\in\mathbb R^{p\times(L+1)}$ and $\gamma\in\Delta_{L+1}$.}
\]
The induced future coefficient is $(B+E)\gamma$. For fixed $(\beta,\gamma)$, GUARD restricts $E$ through  
\begin{equation}
\mathcal E_{\beta,\gamma}(V)
=
\left\{
E\in\mathbb R^{p\times(L+1)}:
E\trans\Sigma_0E
\preceq
\frac{\kappa_0^2\gamma\trans V\gamma}
     {d_{\beta}^2}V
\right\}, \quad \mbox{where}\ d_{\beta} = 2\|\beta-\beta_0\|_{\Sigma_0}.
\label{eq:E_ambiguity_class}
\end{equation}
The factor $\gamma^T V \gamma$ calibrates the perturbation size to the uncertainty of the selected mixture direction, while $d_\beta$ places the perturbation on the same scale as the excess-risk contrast. Let 
\begin{equation}
\mathcal P_{\beta,\gamma}^{\mathrm{pert}}(V)
=
\left\{
\Pbb_{\gamma,E}:
E\in\mathcal E_{\beta,\gamma}(V),\
\Qbb_{\varepsilon_\ell,Y\mid X}^{(\ell)}\text{ satisfies }\eqref{eq: Ql perturb def}\text{ for all } \ell\in[L+1]
\right\}
\label{eq:P_ambiguity_class}
\end{equation}
denote the class of perturbed future populations generated by convex mixtures of locally perturbed target and source conditional laws. We define the population GUARD update as:
\begin{definition}[Population GUARD]
For $\kappa_0>0$ and $V=D^TV_0D\in\mathbb S_+^{L+1}$, the GUARD coefficient is any solution to
\begin{equation}
\beta\subGUARD^\star
\in
\operatorname*{arg\,min}_{\beta\in\mathbb R^p}
\;
\sup_{\gamma\in\Delta_{L+1}}
\;
\inf_{\Pbb\in\mathcal P_{\beta,\gamma}^{\mathrm{pert}}(V)}
\Ebb_\Pbb\bigl\{h_\beta(X,Y)\bigr\}.
\tag{GUARD$_{\scriptscriptstyle \sf DRO}$}
\label{eq:guard_dro_primary}
\end{equation}
\end{definition}
\vspace{-.1in}

\noindent The outer supremum ranges over possible future mixtures of the observed conditional mechanisms, while the inner infimum applies an uncertainty-calibrated adjustment to each mixture. Thus, mixtures driven by poorly estimated source directions are discounted relative to similarly target-compatible but more reliably estimated mixtures.

\begin{proposition}
For fixed $\beta\in\mathbb R^p$ and $\gamma\in\Delta_{L+1}$, the lower-envelope value induced by \eqref{eq:E_ambiguity_class}--\eqref{eq:P_ambiguity_class} satisfies
\begin{equation}
\inf_{\Pbb\in\mathcal P_{\beta,\gamma}^{\mathrm{pert}}(V)}
\Ebb_\Pbb\bigl\{h_\beta(X,Y)\bigr\}
=
\Ebb_{\Pnew(\gamma)}
\bigl\{h_\beta(X,Y)\bigr\}
-
\kappa_0\gamma\trans V\gamma .
\label{eq:lower_envelope_identity}
\end{equation}
Consequently, the GUARD target in \eqref{eq:guard_dro_primary} equivalently solves
\begin{equation}
\beta\subGUARD^\star
\in
\operatorname*{arg\,min}_{\beta\in\mathbb R^p}
\;
\sup_{\gamma\in\Delta_{L+1}}
\left[
\Ebb_{\Pnew(\gamma)}
\bigl\{
(Y-A\trans\beta)^2-(Y-A\trans\beta_0)^2
\bigr\}
-
\kappa_0\gamma\trans V\gamma
\right].
\label{eq:guard_variance_adjusted}
\end{equation}    \label{prop1}
\end{proposition}
\vspace{-.2in}

\noindent This form makes the role of $V$ explicit. GUARD adjusts the target-guided excess-loss by the uncertainty of each mixture direction, encouraging source borrowing  only when the mixture is both target-compatible and precisely estimated. 

\subsection{Transfer Learning Interpretation}
\label{subsec:guard_transferability}
The GUARD objective also has equivalent source-mixture interpretations. Let $B^\dagger(\kappa_0)$ denote a randomly perturbed coefficient library centered at $B$, satisfying
\[
\Ebb[B^\dagger(\kappa_0)]=B,
\qquad
\Ebb\left\{
[B^\dagger(\kappa_0)-B]\trans\Sigma_0[B^\dagger(\kappa_0)-B]
\right\}
=
\kappa_0 V .
\]

\begin{theorem}
\label{thm:guard_identification}
Assume that $\Sigma_0$ and $V$ are positive definite. The GUARD coefficient of Definition~\ref{eq:guard_dro_primary} can be written as
$\beta^\star\subGUARD=B\gamma^\star$, where
\begin{equation}
\tag{GUARD$_{\scriptscriptstyle \sf Pen}$} 
\label{eq:guard_penalization_mixture}
\gamma^\star=\arg\min_{\gamma\in\Delta_{L+1}}
\bigl\{\gamma^{\!\top}V\gamma+\lambda\,\|B\gamma-\beta_0\|_{\Sigma_0}^2\bigr\}.
\end{equation}
The same $\gamma^\star$ also admits two equivalent characterizations:
\begin{align}
\gamma^\star & =
\operatorname*{arg\,min}_{\gamma\in\mathcal{G}_{\tau}} \gamma^TV\gamma \text{ with }\mathcal G_{\tau} = \left\{
\gamma\in\Delta_{L+1}:
\|B\gamma-\beta_0\|_{\Sigma_0}^2\leq \tau
\right\};
\tag{GUARD$_{\scriptscriptstyle \sf Cons}$} 
\label{eq:guard_constraint_mixture} \\
\gamma^\star
& =
\operatorname*{arg\,min}_{\gamma\in\Delta_{L+1}}
\Ebb_{B^\dagger(\kappa_0)}
\left[
\left\|
B^\dagger(\kappa_0)\gamma-\beta_0
\right\|_{\Sigma_0}^2
\right].
\tag{GUARD$_{\scriptscriptstyle \sf Exp}$} 
\label{eq:guard_expectation_mixture} 
\end{align}
When the constraint in \eqref{eq:guard_constraint_mixture} is active, the correspoding multiplier in \eqref{eq:guard_penalization_mixture} satisfies
\begin{equation}
\lambda(\tau) = \frac{f_V(\gamma^*(\tau))}{\tau} \quad \text{where} \quad f_V(\gamma):=(e_0-\gamma)\trans V\gamma.
\label{eq: lambda-tau relationship}
\end{equation}
\end{theorem}

\begin{figure}[H]
    \centering
    \includegraphics[width=0.5\linewidth]{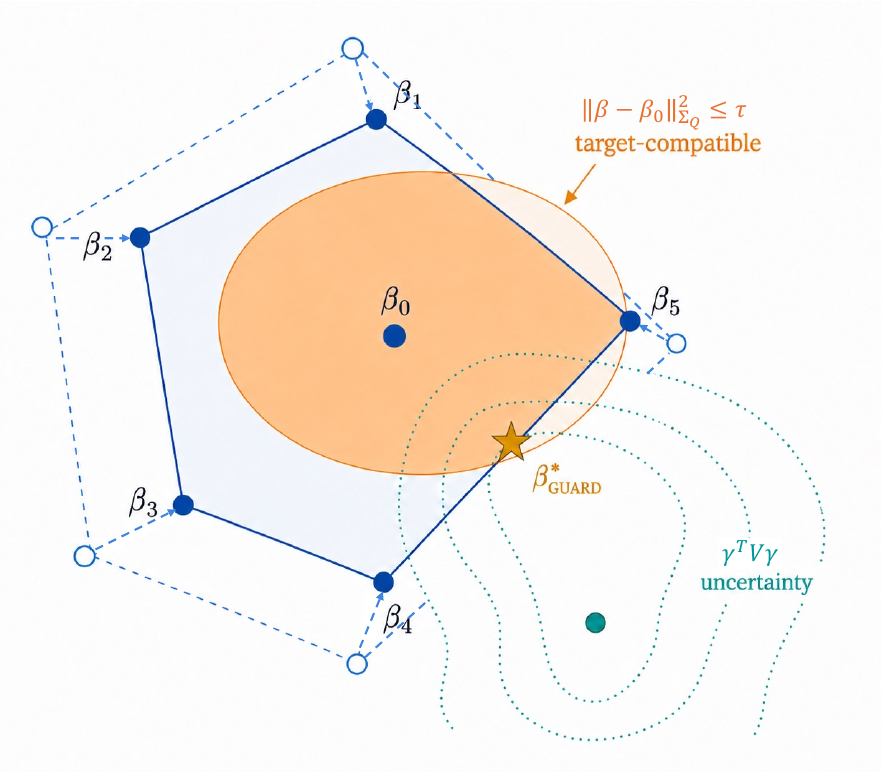}
    \caption{Illustration of \eqref{eq:guard_constraint_mixture}}
    \label{fig: guard}
\end{figure}

Representation \eqref{eq:guard_expectation_mixture} gives a Bayesian
decision view. It treats the coefficient library $B$ as uncertain and selects
the mixture weight $\gamma$ by minimizing the averaged distance between the
induced update $B^\dagger(\kappa_0)\gamma$ and the target benchmark
$\beta_0$. Thus, the action is the source-mixture weight, and the loss is
averaged over uncertainty in the target and source coefficient library.
Representation \eqref{eq:guard_constraint_mixture} gives the transfer-learning
view, as also illustrated in Figure \ref{fig: guard}. The set $\mathcal G_\tau$ contains mixtures whose induced coefficient
$B\gamma$ remains close to the target update $\beta_0$. Within this set,
GUARD chooses the mixture with the smallest uncertainty $\gamma\trans V\gamma$.
This defines transferability at the mixture level rather than at the
individual-source level, allowing heterogeneous sources to contribute jointly
even when no single source is close to the target.
Representation \eqref{eq:guard_penalization_mixture} gives the computational
view. The parameter $\lambda(\tau)$ controls the tradeoff between proximity to
$\beta_0$ and uncertainty of the selected mixture, making this form convenient
for estimation and tuning.

\section{GUARD Estimation and Inference}\label{sec:estimation}

We now describe estimation and inference for the GUARD update. We first
present the estimation procedure, which has three components: estimating the
target-projected coefficient library
$B=(\beta_0,\beta_1,\ldots,\beta_L)$, estimating its uncertainty matrix $V$,
and solving the sample GUARD optimization problem to obtain
$\widehat\beta\subGUARD$. We then turn to inference. Because the estimated GUARD weight is the solution to a constrained quadratic program and can be nonsmooth in the estimated inputs, we use a screened Bayesian bootstrap to resample the estimation-and-optimization procedure and form a union
confidence interval (CI) over stable resamples. 

For each population $\ell\in[L]_+$, the observed data available for
estimation are $\Dl=\Ll\cup\Ul$, where
$\Ll=\{(X_i^{(\ell)},Y_i^{(\ell)})\}_{i=1}^{n_\ell}$
contains the labeled observations and
$\Ul=\{X_i^{(\ell)}\}_{i=n_\ell+1}^{n_\ell+N_\ell}$
contains the additional unlabeled covariates. Let
$\PbbhatLl$ and $\PbbhatUl$ denote the empirical measures over
$\Ll$ and $\Ul$, respectively. To implement cross-fitted DML estimators, we split each observed dataset $\Dl$ into two folds, $\Isc_1$ and $\Isc_2$. The labeled and unlabeled empirical measures corresponding to data in fold $k\in \{1,2\}$th are denoted by $\PbbhatLkl$ and $\PbbhatUkl$.
We assume that labels are missing completely at random within each population,
and that $N_\ell\asymp N_0$ for $\ell\in[L]$ with
$N_0\gg \max_{\ell\in[L]_+} n_\ell$.

\subsection{Estimation}
\subsubsection{Estimating the target-projected coefficient library $B$}
\label{subsec:estimation_B}

The estimation of $B$ combines semi-supervised learning and DML. The semi-supervised component uses the large pool of unlabeled target covariates to estimate the target geometry $\Sigma_0=\Et(AA\trans)$ as $\widehat\Sigma_0= \PbbhatUt(AA\trans)$ and to evaluate source mechanisms under the target covariate distribution. For each source $\ell$, we flexibly estimate $m_\ell(x)$ and the density ratio $w_\ell(x)=d\Pt_X(x)/d\Pl_X(x)$, and use a cross-fitted orthogonal moment to reduce first-order nuisance estimation bias, where $d\Pl_X$ denotes the density function of $X\supl$. 

We first use ordinary least square (OLS) with the labeled target data to estimate $\beta_0$ as
\begin{equation}
\widehat\beta_0=[\PbbhatLt(AA\trans)]^{-1}\PbbhatLt(AY).
\label{eq:target_beta_hat_main}
\end{equation}
For source $\ell \in [L]$, recall that
$\beta_\ell=\Sigma_0^{-1}\mu_\ell$, where
$\mu_\ell=\Et\{A m_\ell(X)\}$. A direct plug-in estimator would fit
$m_\ell$ in the source and average $A\widehat m_\ell(X)$ over the target
covariates. Following DML strategies, to reduce sensitivity to errors in $\widehat m_\ell$, we use the orthogonal identity   
\[
\mu_\ell \equiv \Et\{A m_\ell(X)\} = \Et\{A m_\ell(X)\} + \El\{w_\ell(X)A(Y-m_\ell(X))\}.
\]
The first term places the source outcome mechanism on the target covariate scale. The second term has mean zero at the truth, but corrects first-order errors in estimating $m_\ell$, with $w_\ell$ adjusting for covariate shift between the source and target.

Using two-fold cross-fitting for notation, let 
$(\widehat m\sublk,\widehat w\sublk)$ be trained on fold $\Isc_k$ and evaluated
on fold $\Isc_{k'}$ for $k\ne k'$. Define
\[
\widehat\mu_\ell^{k,k'}
=
\PbbhatUkt\{A\widehat m_{\ell,k'}(X)\}
+
\PbbhatLkl[
\widehat w_{\ell,k'}(X)A\{Y-\widehat m_{\ell,k'}(X)\}].
\]
We set
\begin{equation}
\widehat\mu_\ell = \frac{1}{2}(\widehat\mu_\ell^{1,2}+\widehat\mu_\ell^{2,1}), \qquad
\widehat\beta_\ell=\widehat\Sigma_0^{-1}\widehat\mu_\ell,
\quad \ell\in[L].
\label{eq:source_beta_hat_main}
\end{equation}
The estimated coefficient library is
\[
\widehat B=(\widehat\beta_0,\widehat\beta_1,\ldots,\widehat\beta_L).
\]

\begin{proposition}
\label{propo:Doubly}
For $\ell \in [L]$, $\widehat \beta_l$ has the convergence rate
\[
\|\widehat{\beta}_\ell-\beta_\ell\|_2
=
O_P\left[
\sqrt{\frac{p}{n_\ell}} + \Delta_{mw}^{(l)}
\right].
\]
where $\Delta_{mw}^{(l)}:=\|\widehat m_\ell-m_\ell\|_{L_2(\Pl_X)}
\,
\|\widehat w_\ell-w_\ell\|_{L_2(\Pl_X)}$. 
\end{proposition}
The leading term is the oracle sampling error, while nuisance estimation enters only through the product $\Delta_{mw}^{(\ell)}$. Thus, flexible methods can be used for both the source outcome regression $m_\ell$ and the density ratio $w_\ell$ without introducing first-order bias in the target-projected source coefficient.

\subsubsection{Estimation of $V$}
\label{subsec:estimate_V}
Given $\Bhat$, we next move to estimate 
$V := \Ebb[(\widehat B-B)\trans\Sigma_0(\widehat B-B)]$ with  $V_{\ell k} := \Ebb [(\widehat\beta_\ell -\beta_\ell)\trans\Sigma_0(\widehat\beta_k-\beta_k)].$
The uncertainty of $\Bhat\gamma$ is $\gamma\trans V\gamma = \Ebb \left[\|\widehat B\gamma-B\gamma\|_{\Sigma_0}^2\right]$. The error in $\betahat_\ell$ has two sources. One comes from the target unlabeled covariates, which are used to estimate the target projection and are also shared across all columns of $\Bhat$, creating correlation between $\betahat_\ell$ and $\betahat_k$. The other comes from the labeled data used to learn each population-specific outcome distribution, which mainly contributes to the diagnal entries of $V$. 

We now define the components that separate these two sources of variation.
For $\ell,\ell'\in[L]_+$, let
\begin{equation}
\begin{aligned}
\Delta_\ell(X)
&=\{A m_\ell(X)-\mu_\ell\}-(AA\trans-\Sigma_0)\beta_\ell,\\
\theta_{\ell\ell'} 
&:=\Et\{\Delta_\ell(X)\trans\Sigma_0^{-1}\Delta_{\ell'}(X)\}, \, \theta_{0\ell}:=0 \qquad \ell,\ell'\in[L]\\
\theta_\ell
&:=\El\{w_\ell(X)^2(Y-m_\ell(X))^2A\trans\Sigma_0^{-1}A\},\qquad \ell\in[L],\\
\theta_0
&:=\Et\{(Y-A\trans\beta_0)^2A\trans\Sigma_0^{-1}A\}
\end{aligned}
\label{eq:theta_exact}
\end{equation}

Here $\Delta_\ell(X)$ is the target-side influence contribution for the
projection of the $\ell$th outcome mechanism. The quantity
$\theta_{\ell\ell'}$ captures the shared target-side covariance between
columns $\ell$ and $\ell'$, induced by the common unlabeled target sample.
The term $\theta_\ell$ is the labeled source contribution to the
variance of $\widehat\beta_\ell$, after reweighting source covariates to the
target distribution. Finally, $\theta_0$ is the labeled target
contribution to the variance of $\widehat\beta_0$. Under cross-population independence of the labeled samples and second-order DML remainders, we show in Appendix that

\begin{corollary}
\label{thm:cov}
Assume $\|\widehat m_\ell-m_\ell\|_{L_2(\Pl)}\,\|\widehat w_\ell-w_\ell\|_{L_2(\Pl)}=o\big(N_0^{-1/2}+n_\ell^{-1/2}\big)$. Then
\begin{alignat}{2}
\label{eq:V_exact_lk}
V_{\ell\ell'} &= \mathbf 1\{\ell,\ell'\in[L]\}\frac{1}{N_0}\theta_{\ell\ell'}+
\1\{\ell=\ell'\}\frac{1}{n_\ell}\theta_\ell + o\big(\frac{p}{\sqrt{n_\ell n_{\ell^{\prime}}}}\big) &\quad& \ell,\ell'\in
[L]_{+}, 
\end{alignat}
Consequently,
\begin{equation}
 V=D+R+o_{\mathrm{op}}\!\left(\frac{p}{n_0}\right), \quad \text{where} \quad  D:=\operatorname{diag}\!\left(
      \frac{\theta_0}{n_0},\frac{\theta_1}{n_1},\ldots,
      \frac{\theta_L}{n_L}\right),
 \quad
 R:=\left[\frac{\theta_{\ell \ell'}}{N_0}\right]_{\ell,\ell'\in[L]_+},
 \label{eq:cor1-V-matrix}
\end{equation}
where the matrix remainder is measured in operator norm.
\end{corollary}
The first term in \eqref{eq:V_exact_lk} explains why $V$ is generally non-diagonal: all the source columns of $\widehat B$ use the same unlabeled target sample. The remaining terms come from mutually independent labeled samples and therefore enter only the diagonal.

We estimate the components in \eqref{eq:V_exact_lk} by cross-fitting, using the nuisance estimates from Section~\ref{subsec:estimation_B}. 
We construct the $\Isc_k \to \Isc_{k'}$ ($k \ne k'$) cross-fitted estimators
for $\theta_{\ell\ell'}$ and $\theta_\ell$ as 
$\widehat\theta_{\ell \ell'}^{k,k'}=
\PbbhatUkt(\widehat\Delta\sublkp\trans\widehat\Sigma_{0,k'}^{-1}
\widehat\Delta_{\ell',k'})$, 
\[
\widehat\theta_{\ell}^{k,k'}
= \PbbhatLkl\left[\widehat w\sublkp^2(X)\{Y-\widehat m\sublkp(X)\}^2
A\trans\widehat\Sigma_{0,k'}^{-1}A\right], \ \ell \in [L], \qquad
\widehat\theta_0^{k,k'} = \PbbhatLkt
\left\{(Y-\betahat_{0,k'}\trans A)^2A\trans\widehat\Sigma_0^{-1}A\right\},
\]
where $\widehat\mu\sublkp$, $\widehat\Sigma_{0,k'}$, and
$\widehat\beta\sublkp$ are the corresponding estimators derived using data in $\Isc_{k'}$ and 
\[
\widehat\Delta\sublkp(X)
=
\{A\widehat m\sublkp(X)-\widehat\mu\sublkp\}
-(AA\trans-\widehat\Sigma_{0,k'})\widehat\beta\sublkp.
\]
Averaging the two fold directions gives
$\widehat\theta_{\ell\ell'}
=\frac{1}{2}(\widehat\theta_{\ell\ell'}^{1,2}+\widehat\theta_{\ell\ell'}^{2,1})
$ and $\widehat\theta_\ell=\frac{1}{2}(\widehat\theta_{\ell}^{1,2}+\widehat\theta_{\ell}^{2,1})$, $\ell, \ell' \in [L]_+$
Finally, we assemble $\widehat V$ entrywise as
\begin{equation}
\widehat V_{\ell\ell'}
=
\mathbf 1\{\ell,\ell'\in[L]\}N_0^{-1}\widehat\theta_{\ell\ell'}
+\mathbf 1\{\ell=\ell'=0\}n_0^{-1}\widehat\theta_0
+\mathbf 1\{\ell=\ell'\in[L]\}n_\ell^{-1}\widehat\theta_\ell,
\qquad \ell,\ell'\in[L]_+ .
\label{eq:V_hat_main}
\end{equation}
In computation, we use the symmetrized matrix
$(\widehat V+\widehat V\trans)/2$, with negative eigenvalues truncated at zero if needed. When nuisance-estimation bias in the variance components is not negligible at the scale required for inference, an additional bias correction is needed. The Supplementary Materials provide a bias-corrected estimator for $V$ and the conditions under which this correction can be omitted.

\subsubsection{Estimation of GUARD}
\label{subsec:sample_guard_optimization}
We estimate GUARD using the penalized formulation in \eqref{eq:guard_penalization_mixture}, which is computationally convenient and directly controls the tradeoff between target
compatibility and uncertainty-aware source borrowing. Given $\widehat B$, $\Vhat$, we define the GUARD estimator $\widehat\beta\subGUARD := \widehat B\widehat\gamma$ with the sample GUARD weight 
\begin{equation}
\begin{aligned}
\widehat\gamma
&=
\arg\min_{\gamma\in\Delta_{L+1}}
\gamma\trans\widehat V\gamma
+
\lambda(\widehat{B}\gamma-\widehat{\beta}_0)\trans
\widehat\Sigma_0
(\widehat{B}\gamma-\widehat{\beta}_0), \\
&=
\arg\min_{\gamma\in\Delta_{L+1}}
\gamma\trans\widehat V\gamma
+
\lambda(\gamma-e_0)\trans
\widehat\Gamma
(\gamma-e_0)
\end{aligned}
\label{eq:guard-sample-pen}
\end{equation}
where $e_0=(1,0,\ldots,0)\trans$ and $\widehat{\Gamma} = \widehat{B}^{\top}\widehat{\Sigma}_0\widehat{B}$. The complete GUARD estimation procedure is summarized in Algorithm \ref{alg:guard_dml}.

\begin{algorithm}[h]
\caption{GUARD estimation with cross-fitted DML source learning}
\label{alg:guard_dml}
\begin{algorithmic}[1]
\Require Target data $\Dsc\supzero=\Lsc\supzero\cup\Usc\supzero$, with $\Lsc\supzero=\{(X_i\supzero,Y_i\supzero)\}_{i=1}^{n_0}$ and $\Usc\supzero=\{X_i\supzero\}_{i=n_0+1}^{n_0+N_0}$; source data $\Dl=\Ll\cup\Ul$ for $\ell\in[L]$; penalty level $\lambda\ge 0$.
\Ensure GUARD estimator $\widehat\beta\subGUARD$, weight $\widehat\gamma$, coefficient library $\widehat B$, and $(\widehat\Sigma_0,\widehat\Gamma,\widehat V)$.
\State Compute $\widehat\Sigma_0=\PbbhatUt(AA\trans)$ and $\widehat\beta_0=[\PbbhatLt(AA\trans)]^{-1}\PbbhatLt(AY)$ as in \eqref{eq:target_beta_hat_main}.
\State Split each $\Dl$, $\ell\in[L]_+$, into two folds $\Isc_1$ and $\Isc_2$. For $k\in\{1,2\}$, let $\PbbhatLkl$ and $\PbbhatUkl$ denote the fold-$k$ labeled and unlabeled empirical measures and let $\widehat\Sigma_{0,k}=\PbbhatUkt(AA\trans)$.
\For{$\ell=1,\dots,L$}
  \For{$k=1,2$}
    \State Fit the outcome regression $\widehat m\sublk$ on the labeled source observations $\Ll\cap\Isc_k$, and the density ratio $\widehat w\sublk$ on the source and target covariates in $(\Dl\cup\Usc\supzero)\cap\Isc_k$.
  \EndFor
  \State For $(k,k')\in\{(1,2),(2,1)\}$, evaluate the nuisance functions trained on $\Isc_{k'}$ on the opposite fold $\Isc_k$:
  \[
  \widehat\mu_\ell^{k,k'}
  =\PbbhatUkt\{A\widehat m\sublkp(X)\}
  +\PbbhatLkl\big[\widehat w\sublkp(X)A\{Y-\widehat m\sublkp(X)\}\big].
  \]
  \State Set $\widehat\mu_\ell=\tfrac12(\widehat\mu_\ell^{1,2}+\widehat\mu_\ell^{2,1})$ and $\widehat\beta_\ell=\widehat\Sigma_0^{-1}\widehat\mu_\ell$ as in \eqref{eq:source_beta_hat_main}.
\EndFor
\State Form $\widehat B=(\widehat\beta_0,\widehat\beta_1,\ldots,\widehat\beta_L)$, $\widehat\Gamma=\widehat B\trans\widehat\Sigma_0\widehat B$, and assemble $\widehat V=[\widehat V_{\ell\ell'}]_{\ell,\ell'\in[L]_+}$ by \eqref{eq:V_hat_main}.
\State Solve the quadratic program \eqref{eq:guard-sample-pen},
\[
\widehat\gamma
=\argmin_{\gamma\in\Delta_{L+1}}\
\gamma\trans\widehat V\gamma+\lambda(\gamma-e_0)\trans\widehat\Gamma(\gamma-e_0),
\qquad e_0=(1,0,\ldots,0)\trans.
\]
\State \Return $\widehat\beta\subGUARD=\widehat B\,\widehat\gamma$, together with $\widehat\gamma$, $\widehat B$, and $(\widehat\Sigma_0,\widehat\Gamma,\widehat V)$.
\end{algorithmic}
\end{algorithm}


\subsection{Perturbation inference for GUARD}
\label{sec:guard-screened-bb-inference}
We next describe inference for a fixed linear contrast of the population GUARD coefficient 
\[
\eta\subGUARD(a):=a\trans\beta^*,
\quad
\beta\subGUARD^*:=B\gamma^*,
\]
where $a\in \Rbb^p$ is pre-specified and $\gamma^*$ is the population solution of the GUARD optimization problem. The main difficulty for inference is that $\gammahat$ is the solution to a constrained quadratic program and can be a non-smooth function of the estimated inputs $(\Bhat, \Vhat, \widehat\Gamma)$ due to the boundary constraint. We therefore use a screened Bayesian-bootstrap procedure. The bootstrap resamples the empirical estimating equations, resolves the GUARD optimization problem, removes resamples with unusually large first-order objective perturbations, and forms a union CI over the remaining stable resamples.

\subsubsection{Perturbation weights}
For each resample $m=1,\ldots,M$, independently generate random weights within each independent data pool, $\{\Dsc\supl=\Lsc\supl\cup \Usc\supl, \ell \in [L]_+\}$. For $\ell \in [L]_+$, draw 
$\{\omega\suplm_i,i=1,...,n_\ell,n_{\ell}+1,...,n_{\ell}+N_{\ell}\}$ independently from $\operatorname{Dirichlet}(1,\ldots,1)$. Let $\PbbhatLlm$ and $\PbbhatUlm$ denote the corresponding weighted empirical operators. Fold specific weighted operators are denoted by $\PbbhatLklm$ and $\PbbhatUklm$, with weights normalized within each fixed fold. 

\begin{remark}
Throughout the resampling procedure, all nuisance estimators obtained in the original estimation step are kept fixed. In particular, $\widehat m\sublk,\widehat w\sublk$ for $l\in [L]$ and $k=1,2$ are not re-fit in any bootstrap replicate. The sample-splitting indices are also fixed. The perturbation only reweights the empirical averages appearing in the cross-fitted estimating equations; it does not re-estimate nuisance functions or redraw sample splits.
\end{remark}

Using the weighted empirical measures, define the resampled target covariance and $\beta_0$ as 
\[
\widehat\Sigma_0\supm=\PbbhatUtm(AA\trans),\qquad
\widehat\beta_0\supm=\widehat\beta_0 + (\widehat\Sigma_{0})^{-1}
\PbbhatLtm\{A(Y - A\trans\widehat\beta_0)\} .
\]
For each source $\ell\in [L]$, the perturbed version of the two-fold
DML estimator is obtained by replacing the empirical measures in the original
estimator with their weighted counterparts:
\[
\widehat\beta_\ell\supm=\frac12
\left(\widehat\beta_\ell^{1,2[m]}+\widehat\beta_\ell^{2,1[m]}\right),
\quad \betahat_\ell^{k,k'[m]} = (\widehat\Sigma_0\supm)^{-1} \widehat\mu_\ell^{k,k'[m]}
\]
with 
\[
\widehat\mu_\ell^{k,k'[m]}
= \PbbhatUktm\{A\widehat m_{\ell,k'}(X)\} + \PbbhatLklm[
\widehat w_{\ell,k'}(X)A\{Y-\widehat m_{\ell,k'}(X)\}].
\]
Collect the resampled columns as $\widehat B\supm
:=
\left[
\widehat\beta_0\supm,
\widehat\beta_1\supm,
\ldots,
\widehat\beta_L\supm
\right]$. Accordingly, the resampled penalty matrix is $\widehat\Gamma\supm
:=
(\widehat B\supm)\trans
\widehat\Sigma_0\supm
\widehat B\supm$.

The resampled variance matrix $\widehat V\supm$ is constructed by applying the same weights to the debiased variance estimator. Specifically, we define
$$
\widehat V_{\ell\ell'}\supm
=
N_0^{-1}\widehat\theta_{\ell\ell'}\supm
+\mathbf 1\{\ell=\ell'=0\}n_0^{-1}\widehat\theta_0\supm
+\mathbf 1\{\ell=\ell'\in[L]\}n_\ell^{-1}\widehat\theta_\ell\supm,
\qquad \ell,\ell'\in[L]_+ .
$$
where $\widehat\theta_{\ell\ell'}\supm$ and $\widehat\theta_\ell\supm$ are obtained from the exact same debiased estimating equations as $\widehat\theta_{\ell\ell'}$ and $\widehat\theta_\ell$, except that each empirical operator is replaced by its weighted version. 

With $\Sigmahat_0\supm$, $\Vhat\supm = [\Vhat_{\ell\ell'}\supm]$, and $\Bhat\supm$, we compute 
the resampled GUARD weight
\[
\widehat\gamma\supm
\in
\argmin_{\gamma\in\Delta}
\left\{
\gamma\trans\widehat V\supm\gamma
+
\lambda(\gamma-e_0)\trans
\widehat\Gamma\supm
(\gamma-e_0)
\right\}, \quad \mbox{where}\  
\Gammahat\supm = (\Bhat\supm)\trans\Sigmahat_0\supm(\Bhat\supm).
\]

\subsubsection{Screening resamples and union confidence interval}
\label{sec:screening-statistic-covariance}
The key difficulty for inference is that GUARD weight $\gamma^*$ is defined as the optimizer of a constrained quadratic program. Small perturbations of $(V,\Gamma)$ can change the estimated weights discontinuously and thus standard asymptotic normality based inference fails to characterize the distribution of $\widehat\eta\subGUARD(a) = a\trans \Bhat \gammahat$ and hence provide valid CI for $\eta\subGUARD(a)$. We thus separate the two sources of unceratin in $\widehat\eta\subGUARD(a)$, $\Bhat$ and $\gammahat$. We first generate a collection of candidate weights $\widehat\gamma^{[m]}$ that are plausible sampling perturbations of $\gamma^*$. Then, for each retrained candidate weight $\gamma_m=\widehat\gamma^{[m]} $, we treat the weight as fixed and construct the usual asymptotic CI for $a\trans B\gamma_m$ based on the distribution of $a\trans \Bhat\gamma_m$. The final confidence set is the union of these fixed-weight intervals.

The screening step removes resamples whose perturbation of the GUARD objective is too
large to be regarded as ordinary sampling variation. To see which quantity should be
screened, note that $\widehat\gamma\supm$ and $\gamma^*$ solve the same quadratic program
with different inputs, $(\widehat V\supm,\widehat\Gamma\supm)$ versus $(V,\Gamma)$. To
first order, the gap between the two solutions is driven by the perturbation of the
objective's gradient at $\gamma^*$,
\[
G_{\tot}\supm := (\widehat V\supm - V)\gamma^* + \lambda(\widehat\Gamma\supm-\Gamma)\delta^*,
\qquad \delta^* := \gamma^* - e_0,
\]
in the sense that $\|\widehat\gamma\supm-\gamma^*\|_2$ is bounded by a multiple of
$\|G_{\tot}\supm\|_2$. Because $\widehat V\supm$ and
$\widehat\Gamma\supm$ are Bayesian-bootstrap perturbations of $\widehat V$ and
$\widehat\Gamma$, this perturbation splits into two parts,
\[
G_{\tot}\supm = G + U_*\supm,\qquad
G := (\widehat V - V)\gamma^* + \lambda(\widehat\Gamma-\Gamma)\delta^*,\quad
U_*\supm := (\widehat V\supm - \widehat V)\gamma^* + \lambda(\widehat\Gamma\supm-\widehat\Gamma)\delta^*.
\]
The first part $G$ is the estimation error of the original sample relative to the population, whereas $U_*\supm$ is the additional perturbation introduced by the $m$th set of weights, and is the only part that varies with $m$. Conditional on the observed
data, $G$ is fixed, so the distance from $\widehat\gamma\supm$ to $\gamma^*$ can be
controlled only through $U_*\supm$. An ideal screening rule therefore retains the replicates for which $U_*\supm$, normalized by its conditional standard deviation, is not unusually large.
Since $U_*\supm$ depends on the unknown $\gamma^*$, we use the plug-in analogue
\[
\widehat U\supm =
(\widehat V\supm-\widehat V)\widehat\gamma +
\lambda(\widehat\Gamma\supm-\widehat\Gamma)(\widehat\gamma-e_0).
\]
We derive the covariance of $U_*\supm$ and its estimator, $\widehat\Omega\subGUARD$, from the conditional randomness of the Bayesian-bootstrap weights in the Supplementary Materials. Let
$\widehat s_j^2:=(\widehat\Omega_{\GUARD})_{jj}$ and  $c_{\alpha}:=z_{1-\alpha/\{2(L+1)\}}$. We define the 
screened set of resamples by 
\begin{equation}
\label{eq:guard-screened-set}
{\mathcal M}_{\alpha_1} := \left\{ 1\le m\le M: \max_{0\le j\le L}
\left|\frac{\widehat U_j\supm}{\widehat s_j}
\right| \le c_{\alpha_1}
\right\},
\end{equation}
where $\alpha_1$ is the allocated type 1 error for selecting plaucible set for $\gamma$. We allocate the overall error level as $\alpha_0 = alpha_1+\alpha_2$, where $\alpha_2$ is used for the fixed-weight confidence intervals. Equivalently, the retained candidate weights are $\{\widehat\gamma^{[m]}:m\in\mathcal M_{\alpha_1}\}$. 

For each retained $m\in\mathcal M$, we now treat $\widehat\gamma^{[m]}$ as a fixed candidate weight and conduct inference for $a\trans B\widehat\gamma^{[m]}$. The plug-in estimate is
\[
\widehat\eta^{[m]}(a):=a\trans\widehat B\,\widehat\gamma^{[m]} .
\]
Let $\widehat V(a)$ be the estimated the covariance matrix of
$(a\trans\widehat\beta_0,\ldots,a\trans\widehat\beta_L)\trans$. Then for any given 
$\widehat{\gamma}\supm$, the estimated variance for $\widehat\eta\supm(a)$ is $(\widehat{\gamma}\supm)\trans\widehat V(a)\widehat{\gamma}\supm$. The resulting $(1-\alpha_2)$-level CI for $a\trans B\widehat{\gamma}\supm$ is 
\[
CI_{\alpha_2}\supm(a \mid \widehat{\gamma}\supm)
:=
\left[
\widehat\eta^{[m]}(a) -
z_{1-\alpha_2/2} \sqrt{(\widehat{\gamma}\supm)\trans\widehat V(a)\widehat{\gamma}\supm} \ ,
\widehat\eta^{[m]}(a) +
z_{1-\alpha_2/2} \sqrt{(\widehat{\gamma)\supm}\trans\widehat V(a)\widehat{\gamma}\supm} 
\right].
\]
The final $(1-\alpha_0)$-level screened union CI for $\eta(a)$ is
\[
CI_{\alpha_0}(a):=
\bigcup_{m\in{\mathcal M}_{\alpha_1}}CI_{\alpha_2}(a \mid \widehat\gamma\supm) , \quad \alpha_0 = \alpha_1+\alpha_2 .
\]

\section{Theoretical Properties}\label{sec:theory}
\subsection{Model Assumptions}
We allow the working-feature dimension $p$ to increase while keeping the number of populations $L_+=L+1$ fixed. Without loss of generality, we assume that the target population has the smallest number of labels, $n_0=\min_{j\in[L_+]}n_j$, and let $n_\star=\max_{j\in[L_+]}n_j$. We organize the theoretical results into three parts: population transferability, estimation error, and inference validity. We state the main assumptions needed for these results below, with additional technical conditions and derivations deferred to the Supplementary Materials.

\begin{assumption}[Sampling, moments, and overlap]
\label{cond:setup}
\label{mt:A1}
The labeled and unlabeled pools are independent, with independent identically
distributed observations within each pool. The sample sizes satisfy
\begin{equation}
 N_j\asymp N_0\ (j\ge1),\quad N_0/n_\star\to\infty,\quad p^2/n_0\to0.
 \label{mt:sizes}
\end{equation}
For fixed positive constants,
\begin{gather}
 \Sigma_0\succeq c_\Sigma I_p,\qquad E_0Y^2\le C_Y,\qquad
 c_w\le w_j(X)\le C_w\quad(j\ge1),\label{mt:model}\\
 \sup_{\|u\|_2=1}\|u^\top A\|_{\psi_2,0}
 +\max_{j\ge1}\|m_j\|_{\psi_2,0}
 +\max_{j\ge1}\|Y-m_j(X)\|_{\psi_2,j}
 +\|Y-A^\top\beta_0\|_{\psi_2,0}\le C.\label{mt:tails}
\end{gather}
The conditional second moments of $Y-A^\top\beta_0$ in the target and
$Y-m_j(X)$ in each source belong to $[c_\epsilon,C_\epsilon]$ almost surely.
For $j\ge1$, $E_j[\{Y-m_j(X)\}^4\mid X]\le C_4$ almost surely.
\end{assumption}

\begin{assumption}[Cross-fitted nuisance accuracy]
\label{mt:A2}\label{cond:nuisance}
Denote $\|f\|_{2,0}=(E_0f^2)^{1/2}$, $q_A=A^\top\Sigma_0^{-1}A$,
$h_j=q_A\{m_j(X)-A^\top\beta_j\}$, and
$v_j=E_j\{q_A(Y-m_j(X))^2\mid X\}$.
On training events with probability tending to one, uniformly over sources
and folds, for $\nu\in\{m,w\}$ and $f=\widehat\nu_j-\nu_j$, 
\begin{equation}
 \left\|\left(1+\frac{q_A}{p}\right)f\right\|_{2,0}
 +\max_{1\le k\le L}\left\|\frac{h_k}{p}f\right\|_{2,0}
 +\sup_{\|u\|_2=1}\|(u^\top A)f\|_{2,0}
 \le \Delta_{\nu,j}.
 \label{mt:L2-mw}
\end{equation}
For the separately trained auxiliary fits,
\begin{equation}
 \left\|\left(1+\frac{q_A}{p}\right)
       \frac{\widehat h_j-h_j}{p}\right\|_{2,0}\le\Delta_{h,j},\qquad
 \left\|\left(1+\frac{q_A}{p}\right)
       \frac{\widehat v_j-v_j}{p}\right\|_{2,0}\le\Delta_{v,j}.
 \label{mt:L2-hv}
\end{equation}
Here $\Delta_{m,j}=\Delta_m(n_j)$ and
$\Delta_{w,j}=\Delta_w(N_0\wedge N_j)$.
The fitted density ratios are bounded, and
$\widehat m_j,\widehat h_j/p,\widehat v_j/p$ have uniformly bounded
$\psi_2,\psi_{2/3},\psi_1$ norms under the target covariate law.
Furthermore,
\begin{equation}
 \max_j\sum_{\nu\in\{m,w,h,v\}}\Delta_{\nu,j}\to0,\qquad
 \max_{j\ge1}\left\{\sqrt p(\Delta_{m,j}+\Delta_{w,j})
             +\sqrt{n_j}\Delta_{m,j}\Delta_{w,j}\right\}\to0.
 \label{mt:nuisance}
\end{equation}
\end{assumption}

\begin{assumption}[Remainder control in $V$]\label{mt:A3}
Assume the difference between V and its first-order approximation $D+R$ satisfies,
\begin{equation}
 \max_{0\le j\le L}\left\{
  \frac{\sqrt{n_j}}p\sum_{k=0}^L n_k|(V-D-R)_{jk}|
  +\frac{|\{(V-D-R)\gamma^\star\}_j|}{s_j}
 \right\}\to0.
 \label{mt:exact-relative}
\end{equation}
\end{assumption}

\begin{assumption}[Gradient regularity]
\label{mt:A4}\label{cond:geometry}
Let $R^V_{j,n}(\gamma)$ and $R^\Gamma_{j,n}(\gamma)$ be the gradient
remainder terms with definitions given in Supplements. Set $s_j^2=(\Omega_{\GUARD})_{jj}$ and $\Lambda=\diag(s_0,\ldots,s_L)$ where  $\Omega_{\GUARD}$ is the population first-order covariance of
$U_\star^{[m]}$ evaluated at $\gamma^\star$. For a fixed $c_C>0$,
\begin{equation}
 \lambda_{\min}(\Lambda^{-1}\Omega_{\GUARD}\Lambda^{-1})\ge c_C,\qquad
 R_{G,n}:=\max_j
 \frac{R^V_{j,n}(\gamma^\star)+R^\Gamma_{j,n}(\gamma^\star)}{s_j}\to0.
 \label{mt:score}
\end{equation}
\end{assumption}

\subsection{Population transferability under source mixing}
\label{subsec:transferability-support-expansion}
The constrained characterization in \eqref{eq:guard_constraint_mixture} provides a mixture-level notion of transferability. For a radius $\tau>0$, the set
\begin{equation}
\mathcal G_\tau := \left\{ \gamma\in\Delta_{L+1}:
\|B\gamma-\beta_0\|_{\Sigma_0}^2\le \tau
\right\}
\label{eq: G tau}
\end{equation}
collects the target-compatible mixtures: each $\gamma\in\mathcal G_\tau$ induces a coefficient $B\gamma$ that remains close to the target benchmark $\beta_0$. This perspective allows transferability to be defined at source-mixture level rather than source by source. 

\begin{definition}[Mixture-level transferability]
\label{def:transferability} 
For a threshold $w\in[0,1]$, define the transferable source set $S_0$ and its target-augmented version $C$ by
\begin{equation}
S_0:=\Big\{
j\in\{1,\ldots,L\}:
\sup_{\gamma\in\mathcal G_\tau}\gamma_j\ge w
\Big\}, \quad C:=\{0\}\cup S_0.
\label{eq:S0-transferability}
\end{equation}
The non-transferable source set is
\begin{equation}
S_1:=\{1,\ldots,L\}\setminus S_0 =\Big\{j\in\{1,\ldots,L\}:\sup_{\gamma\in\mathcal G_\tau}\gamma_j<w
\Big\}.\label{eq:S1-nontransferability}
\end{equation}
\end{definition}
A source belongs to $S_0$ if it can receive at least weight $w$ in some
target-compatible mixture, whereas a source belongs to $S_1$ only if its weight remains uniformly below $w$ over all such mixtures. When $w=1$, this reduces to the conventional source-wise condition $\|\beta_j-\beta_0\|_{\Sigma_0}^2\le\tau$. When $w<1$, a source need not be individually close to the target, as long as it can contribute meaningfully to a target-compatible mixture.

We next introduce a mixture-level separation condition that rules out spurious cancellation by non-transferable sources. For any $\gamma_C\in\Delta_C$, let $\mathcal E_\tau(\gamma_C;S_1)$ be the set of target-compatible mixtures obtained by preserving the relative allocation within $C$ while adding mass supported only on $S_1$:
\begin{equation}
\begin{aligned}
\mathcal E_\tau(\gamma_C;S_1) :=
\Big\{
\gamma=\widetilde{\gamma}_C+\delta_{S_1}\in\mathcal G_\tau:\;
\operatorname{supp}(\widetilde{\gamma}_C)\subseteq C,\quad
\operatorname{supp}(\delta_{S_1})\subseteq S_1, \quad
\frac{\widetilde{\gamma}_C}
{\mathbf 1\trans\widetilde{\gamma}_C}=\gamma_C\Big\}.
\end{aligned}
\label{eq:feasible-support-expansion-class}
\end{equation}
Here, $\widetilde{\gamma}_C$ is the component supported on the target and transferable sources, while $\delta_{S_1}$ is a perturbation supported on the non-transferable sources. The normalization keeps the relative allocation within $C$ unchanged. 

\begin{assumption}
There exists a constant
$c_{\mathrm{sep}}$ such that, for every $\gamma_C\in\Delta_C$ satisfying
$\mathcal E_\tau(\gamma_C;S_1)\neq\varnothing$ and
$\|B_C\gamma_C-\beta_0\|_{\Sigma_0}>0$,
\begin{equation}
\inf_{\gamma\in\mathcal E_\tau(\gamma_C;S_1)}
\frac{\|B\gamma-\beta_0\|_{\Sigma_0}}{
\|B_C\gamma_C-\beta_0\|_{\Sigma_0}}
\ge c_{\mathrm{sep}}>0.
\tag{Sep}
\label{eq:sep-support-expansion}
\end{equation}
When $B_C\gamma_C=\beta_0$, the transferable profile already exactly recovers the target benchmark, and no additional non-improvement requirement is needed.
\label{assumption: sep}
\end{assumption}

Condition \eqref{eq:sep-support-expansion} rules out a cancellation shortcut: starting from any profile supported on $C$, adding sources from $S_1$ cannot reduce the distance to $\beta_0$ below a fixed fraction of the original clean-block distance. Thus, sources in $S_1$ are non-transferable not only because their feasible weights are small, but also because their inclusion cannot substantially improve target compatibility.

This perturbation view suggests that the full GUARD solution should be close to a clean-only oracle that ignores the non-transferable block. To formalize this, let $\eta^*:= [(\eta_C^*)^{\top},
\mathbf{0}^{\top}_{S_1}]^{\top}\in\Delta_{L+1}$ where $\eta_C^*$ is the clean-set-only analog of $\gamma^*$ on $C$:
\[\eta_C^* \in \arg\min_{\gamma_C\in\Delta_C}
\gamma_C\trans V_C\gamma_C\quad \text{subject to}
\quad \|B_C\gamma_C-\beta_0\|_{\Sigma_0}^2\le \tau,
\] 
Thus, $\eta^*$ is the oracle variance-minimizing target-compatible mixture that uses only the target and transferable sources. 
\begin{corollary}[Clean-oracle approximation]
\label{cor:clean-oracle-approx}
Define $\lambda_\Delta := \inf_{\delta\neq0:\ \mathbf 1\trans\delta=0}
\frac{\delta\trans V\delta}{\|\delta\|_2^2}$, then under Assumption \ref{assumption: sep}, we have
\[
\|\gamma^*-\eta^*\|_2^2\lesssim \frac{1}{\lambda_\Delta}
\frac{\tau+L^2w^2}{n_0}.\]
\end{corollary}
Corollary~\ref{cor:clean-oracle-approx} shows that GUARD can
approximate the clean-only oracle without explicitly identifying $S_0$. 
The approximation error is controlled by the target-compatible radius $\tau$, the low-participation threshold $w$, and the covariance geometry encoded in $V$.

\subsection{Estimation error bound}
\label{subsec:estimation_error_bound}
We next study the estimation error of the sample GUARD estimator. The first result gives a estimation error bound for $\widehat{\beta}\subGUARD(\lambda)$ with a fixed penalty level $\lambda\geq0$, showing how errors in the estimated inputs $(\widehat V,\widehat\Gamma)$ propagate to the final estimator. 

\begin{theorem}
\label{thm:guard-pen-rate}
Let $r_\Delta:=\operatorname{rank}\{\Sigma_0^{1/2}
(\beta_1-\beta_0,\ldots,\beta_L-\beta_0)\} \leq L_+$. Under Assumptions~\ref{mt:A1}--\ref{mt:A2}, for a fixed $\lambda\ge0$,
\begin{align}
 &\|\widehat\beta_{\GUARD}(\lambda)-\beta_{\GUARD}^\star(\lambda)\|_2
 \nonumber\\[-2pt]
 &\quad=O_P\!\left[
 \left\{p\sum_{j=0}^{L}\frac{(\gamma_{\lambda,j}^\star)^2}{n_j}\right\}^{1/2}
 +\min\left\{\frac{\lambda n_\star}{p}(1-\gamma_{\lambda,0}^\star),1\right\}
       \sqrt{\frac{r_\Delta}{n_0}}
 +\min\{\lambda,1\}\sqrt{\frac p{n_0}}+r_{n,\lambda}\right],
 \label{eq:guard-gamma-det-bound}
\end{align}
where $r_{n,\lambda}$ collects higher-order nuisance error and the exact-uncertainty discrepancy, defined in the supplements.
\end{theorem}

Theorem \ref{thm:guard-pen-rate} establishes estimation validity for a fixed $\lambda$. To connect this result to  transferabilty, we use the constrained characterization in \eqref{eq:guard_constraint_mixture}, where $\tau$  controls the target-compatible set $\mathcal{G}_\tau$ in \eqref{eq: G tau}.
When the contraint is active, $\tau$ and $\lambda$ are linked through the KKT multiplier $\lambda(\tau)$. We next show that under the clean source-mixing and separation, an appropriate calibration of $\tau$ can exploit transferable sources and improve over target-only estimation.

To make the optimal rate explicit under this special case, we first define source-only clean mixture approximation error
\[
\alpha
:=
\min_{\eta\in\Delta_{S_0}}
\|B_{S_0}\eta-\beta_0\|_{\Sigma_0}^2
\]
Among source mixtures that attain this optimal approximation level, let
\[
\eta_{\base}
\in
\operatorname*{arg\,min}_{\eta\in\Delta_{S_0}}
\eta\trans D_{S_0S_0}\eta
\quad
\text{subject to}
\quad
\|B_{S_0}\eta-\beta_0\|_{\Sigma_0}^2\le \alpha.
\]
Thus, $\eta_{\base}$ attains the best clean-source approximation to $\beta_0$, and among such mixtures, minimizes the leading labeled-sample variance component.   The corresponding effective sample size is
\[
n_{\mix} := n_0+n_{S_0}, \quad \text{where} \quad n_{S_0}:=
\left(\eta_{\base}\trans D_{S_0S_0}\eta_{\base}\right)^{-1}.
\]
Here $n_{S_0}$ represents the effective sample size of the optimal transferable source mixture, and $n_{\mix}$ is the oracle benchmark sample size obtained by combining target information with the optimally weighted clean-source mixture. 
\begin{corollary}
\label{cor:lambda-clean-mixing-main}
Under Assumptions~\ref{mt:A1}--\ref{mt:A2}, there exists a $\lambda^*\ =(n_0/n_{\mathrm{mix}})^{1/3}$ such that
\begin{equation}
 \|\widehat\beta_{\GUARD}(\lambda^*)-\beta_0\|_2
 =O_P\!\left\{\sqrt\alpha+
    \sqrt{\frac{p}{n_{\mathrm{mix}}^{2/3}n_0^{1/3}}}
    +\sqrt{\frac{r_\Delta}{n_0}}+r_n^V\right\}.
    \label{eq:main-target-rate}
\end{equation}
If additionally $n_{\mathrm{mix}}\gtrsim n_0(p/r_\Delta)^{3/2}$, the choice
$\lambda^*=\sqrt{r_\Delta/p}$ gives the bound
\[
 \|\widehat\beta_{\GUARD}(\lambda^*)-\beta_0\|_2 = O_P\{\sqrt{\alpha+p/n_{\mathrm{mix}}}+\sqrt{r_\Delta/n_0}+r_n^V\}.
\]
\end{corollary}

\subsection{Inference validity for screened union CI}
We now justify the screened Bayesian-bootstrap inference procedure for $a\trans\beta^*\subGUARD$. The key
technical issue is the nonsmooth dependence of the GUARD optimizer on
$(V,\Gamma)$. The screening step ensures that the retained bootstrap
perturbations are compatible with ordinary sampling variation, while the union
over retained fixed-weight intervals protects coverage when the optimizer
changes nonsmoothly.

Let $R_{\alpha_1} := c_{\alpha_1}\sqrt{(L+1)C_\Omega}$. For $R<\infty$, define
\[
C_{\mathrm{BB}}(R)
:=
\frac{
\operatorname{Vol}\{B_{L+1}(0,1)\}
}{
2(2\pi)^{(L+1)/2}C_\Omega^{(L+1)/2}
}
\exp\left\{
-\frac{(R+1)^2}{2c_\Omega}
\right\},
\]
where $\operatorname{Vol}\{B_{L+1}(0,1)\}$ is the volume of the unit
Euclidean ball in $\mathbb R^{L+1}$. Define the resampling accuracy
\[
\operatorname{err}_{a_n}(M_n)
:=
\left\{
\frac{\log a_n}
{C_{\mathrm{BB}}(R_{\alpha_1})M_n}
\right\}^{1/(L+1)}.
\]

\begin{theorem}[Existence of a good screened GUARD resample]
\label{thm:guard-good-screened-resample}
Suppose Assumption \ref{mt:A1}--\ref{mt:A4} hold, and the number of bootstrap draws $M_n$ satisfies $\frac{\log M_n}{N_0}\to0$, $\max_{0\le l\le L}\frac{\log M_n}{n_l}\to0$. Also suppose that $\|\widehat\gamma-\gamma^*\|_2=O_p(a_n^{-1/2})$. Then
\[
\liminf_{n\to\infty}
P\left\{
\min_{m\in{\mathcal M}_{\alpha_1}}
\|\widehat\gamma^{[m]}-\gamma^\star\|_2
\le
\frac{
4\,\operatorname{err}_{a_n}(M_n)
}{
c_0\sqrt{a_n}
}
\right\}
\ge
1-\alpha_1.
\]
\end{theorem}

Theorem~\ref{thm:guard-good-screened-resample} is the key link between the screening step and the coverage argument. It shows that, with probability at least $1-\alpha_1$ asymptotically, the screened set $\mathcal M$ contains at least one resample $m^*$ whose optimizer is close to the population GUARD optimizer. This result is sufficient for coverage because the final confidence set is a union over all screened resamples. Therefore, if the interval corresponding to this good resample covers the target, then the whole union interval also covers the target. The coverage theorem below formalizes this argument.

\begin{theorem}[Coverage]
\label{thm:guard-screened-union-coverage}
Suppose the conditions of Theorem~\ref{thm:guard-good-screened-resample} hold. Then the screened union CI $CI_{\alpha_0}(a)$ satisfies
\[
\liminf_{n\to\infty}\liminf_{M\to\infty}
\PP\left\{
a\trans\beta^*\subGUARD
\in
CI_{\alpha_0}(a)
\right\}
\ge
1-\alpha_0.
\]
\end{theorem}

\begin{theorem}[Length]
\label{thm:guard-screened-union-length}
Suppose the assumptions of Theorem~\ref{thm:guard-good-screened-resample} hold. Denote $C_U:=\frac{4c_{\alpha_1}\sqrt{(L+1)C_\Omega}}{c_0}$.
Then there exists some positive constant $C>0$ such that,
\begin{equation}
\lim_{n\to\infty}\lim_{M\to\infty}\PP\left[\operatorname{length}\{CI_{\alpha_0}(a)\}
\leq C\left\{\frac{C_U\|a\|_2}{\sqrt{a_n}}
+\sqrt{\sum_{\ell=0}^L\frac{(\gamma_\ell^\star)^2}{n_\ell}}\right\}\right]=1
\label{eq:length-bound}
\end{equation}
\end{theorem}

\section{Simulation studies}
\label{sec:sim}


\subsection{Simulation settings, benchmarks and evaluation metrics}
\label{sec:sim-setup}
We evaluate GUARD under three source--target configurations. Design~I studies future generalizability under temporal drift, while Designs~II and~III study current-target transfer under heterogeneous and partially corrupted source libraries. Across all designs, we assess point estimation and uncertainty quantification. Across all designs, $X^{(0)}\sim N(0,\Sigma_0)$ with $p=25$, $(\Sigma_0)_{jk}=0.3^{|j-k|}$, and the working design $A=a(X)$ uses the first $p_A=20$ coordinates. Outcomes are generated as
$Y=m_\ell(X)+\varepsilon$, where $\varepsilon\sim N(0,0.5^2)$. The target contains $n_0=200$ labeled and $N_0=10^5$ unlabeled observations, whereas each source contains $n_\ell=2{,}000$ labeled and $N_\ell=N_0$ unlabeled observations, except in Design III where the clean sources have imbalanced labeled sample sizes.
In Designs~I and~II, the sources share the target covariance but have a shifted mean, $X^{(\ell)}\sim N(\mu_S,\Sigma_0)$, with $\mu_S$ equals $0.6$ in the first 10 coordinates and zero elsewhere; in Design~III, all populations share the target covariate distribution. The designs differ in how the source conditional outcome laws relate to the target. Full data-generating details are provided in the Supplementary Materials. \vspace{-.15in}

\paragraph{Design~I: temporal domains and future generalization.} We generate $L=5$ ordered historical source domains and one current target domain to mimic temporal drift within a population. Historical sources gradually depart from a common signal, while the current target is generated from a mixture of recent sources plus a target-specific deviation controlled by $\alpha$, with larger $\alpha$ indicating weaker source-mixture transferability. We then generate post-target future populations under random, perturbed-hull, and adversarial-hull drift to evaluate robustness beyond the observed target. \vspace{-.15in}

\paragraph{Design~II: heterogeneous sources and current-target transfer.} We generate $L=5$ source sites with distinct nonlinear outcome mechanisms and one current target. No single source is directly transferable, but source mixing can approximate the target, with $\alpha$ controlling the degree of target-specific deviation. This design evaluates transfer from heterogeneous sources. \vspace{-.15in}

\paragraph{Design~III: clean and corrupted source libraries.} We generate $L=10$ sources, including nine transferable sources and one corrupted source whose distance from the target is controlled by $\kappa$. In III(i), each clean source matches the target but differs in estimation precision. In III(ii), no clean source matches the target individually, but a mixture of the clean sources recovers the target. Thus, Design~III evaluates both uncertainty-aware borrowing across sources of different precision and robustness to negative transfer from an increasingly corrupted source. \vspace{-.15in}

\paragraph{Benchmarks.} We consider a range of benchmark methods.  \textbf{Target only} fits OLS using only the $n_0$ labeled target observations; while \textbf{pooled} fits OLS after combining all source and target observations. \textbf{TransGLM} \citep{tian2023transglm} is a traditional transfer-learning algorithm that borrows information from sites with similar $\bbeta$. \textbf{Maximin} is the standard group DRO estimate of \citet{meinshausen2015maximin}, with CIs constructed using the resampling method of \citet{guo2023maximin}.  \textbf{RIFL} \citep{guo2025RIFL} is a robust integrative method that assumes a majority of the sites share a common $\bbeta$. It first compares sites pairwise based on their estimated coefficients, uses these pairwise similarities to identify a mutually similar set of sites, and then constructs CIs for the shared $\bbeta$ by applying a resampling-based procedure to the selected sites. We also include \textbf{source--target mixing}, which constructs an aggregate estimator under the source-mixing assumption by choosing simplex weights over a target-source coefficient library. Specifically, it splits the target labels, using the first half to estimate a target column $\hat\bbeta_0^{(1)}$ and the second half to choose simplex weights over $\tilde\bB=(\hat\bbeta_0^{(1)}, \widehat{B}_{\text{source}})$ by least squares. Its inference procedure follows the same resampling idea used for Maximin. \textbf{GUARD} is throughout the version whose $\lambda$ is chosen by Mallows' $C_p$; $\hat m_\ell$ is fit by gradient boosting and $\hat w_\ell$ by cross-fitted logistic regression, and inference uses $M=50$ Bayesian-bootstrap resamples with screening levels $\alpha_1=0.01$, $\alpha_2=0.04$ at nominal $95\%$.

\paragraph{Evaluation Metrics.} For evaluating future generalization under Design I, we consider regret on a future population,
\begin{equation}
\mathrm{Reg}(\hat\bbeta;\bbetafut)
=\Snorm{\hat\bbeta-\bbetafut}-\Snorm{\bbeta_0-\bbetafut}.
\label{eq:regret}
\end{equation}
At each parameter value indexing a future family (i.e., $\delta$ for random drift, $\varepsilon_v$ for perturbed hull, and $r$ at a fixed $\varepsilon_v$ for two adversarial families), we generate 300 future $\bbetafut$, denoted by $\Bscrfut$, and report the worst-case regret $\Rworst = \sup_{\bbetafut \in \Bscrfut} \mathrm{Reg}(\hat\bbeta;\bbetafut)$. For current-target transfer learning under Design II and III, we report estimation error for $\bbeta_0$: $\Snorm{\hat\bbeta-\bbeta_0}$. For inference, we separate three questions. First, we assess validity of the screened GUARD confidence interval for a fixed $\lambda$, reporting coverage and length; these diagnostics are reported in the Appendix. Second, for future generalization under Design I, coverage and length cannot be read separately because a sufficiently wide interval can cover any future $\bbetafut$. We therefore score the two jointly through the Winkler interval score (IS) \citep{winkler1972decision, gneiting2007proper}. 
Third, for current-target transfer under Designs II and III, we report coefficient-level coverage of $\bbeta_0$ and mean interval length, with $\lambda$ perturbed as part of the resampling procedure. Coverage is defined as the fraction of the $p_A$ coordinates whose intervals contain the true coefficient. For level $1-\alpha_0$, an estimated interval $[\widehat{\theta}_\ell,\widehat{\theta}_u]$ and a target $\theta$, the Winkler IS is defined as
\begin{equation}
\mathrm{IS}(\widehat{\theta}_\ell,\widehat{\theta}_u;\theta)
=(\widehat{\theta}_u-\widehat{\theta}_\ell)
+\frac{2}{\alpha_0}\,(\widehat{\theta}_\ell-\theta)\,\1\{\theta<\widehat{\theta}_\ell\}
+\frac{2}{\alpha_0}\,(\theta-\widehat{\theta}_u)\,\1\{\theta>\widehat{\theta}_u\}. 
\label{eq:winkler}
\end{equation}
Lower IS favors intervals that are both short and well calibrated. 

\subsection{Results}
\label{sec:sim-results}

\subsubsection{Estimation and Prediction Performance}
\label{sec:studyA}

\begin{figure}[H]
\centering
\includegraphics[width=1\textwidth]{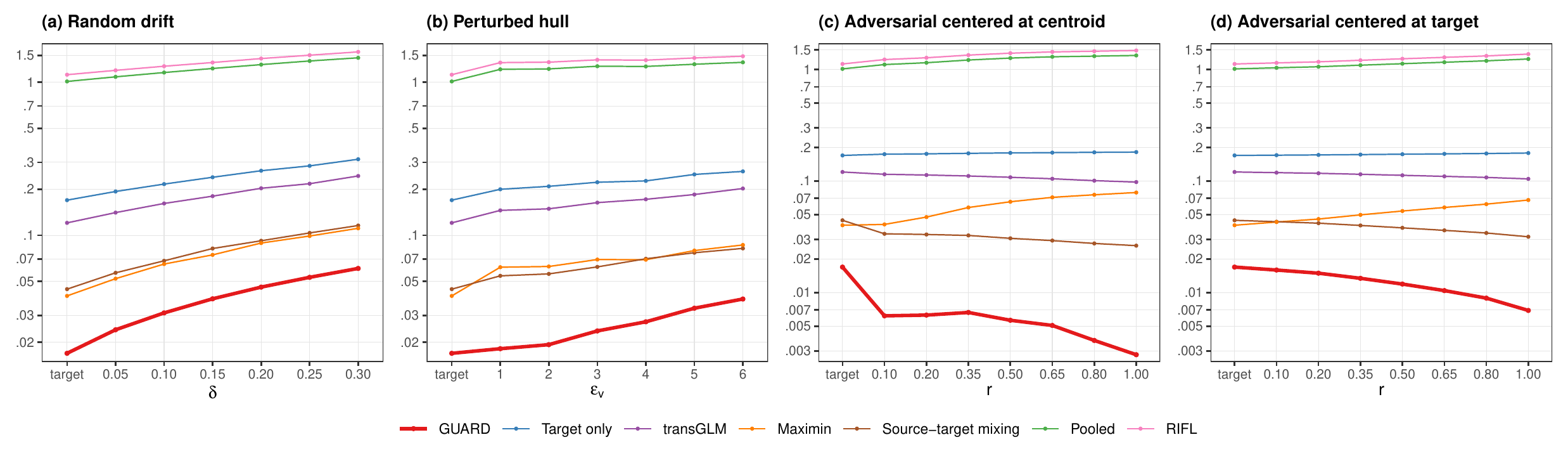}
\caption{\textbf{Future generalization under temporal-domain Design I.} Empirical worst-case regret $\Rworst$ on future populations. Panels correspond to random drift (a); perturbed hull (b); adversarial hull drift with $c=(L+1)^{-1}\bf1$ (c) and with $c=e_0$ (d). The leftmost points on each panel correspond to $\bbetafut=\bbeta_0$, so regret equals current-target  estimation error. 
}
\label{fig:studyA}
\end{figure}

We first evaluate future generalization under Design~I. Figure~\ref{fig:studyA} compares worst-case regret under four future-population generators. When the future population concides with the current target, corresponding to the leftmost points, regret reduces to MSE. GUARD attained an MSE of $0.017$, compared with $0.036$ for source--target mixing, $0.038$ for maximin, $0.11$ for transGLM, and $0.16$ for target only. As the future population moved away from the current target, GUARD retained the most stable. Under random drift (Figure~\ref{fig:studyA}a), GUARD's regret increased from $0.017$ to $0.062$ as $\delta$ increased from 0 to 0.3, more slowly than all other methods. Under the perturbed-hull drift (Figure~\ref{fig:studyA}b), GUARD's regret increased only from $0.022$ at $\varepsilon_v=1$ to $0.046$ at $\varepsilon_v=6$, and remained the lowest over the reported range. These results indicate that GUARD retained its future-generalization advantage as the future population moved either in an unanticipated direction or within a perturbed version of the coefficient library. 

The adversarial-hull results (Figure~\ref{fig:studyA}c,\ref{fig:studyA}d) show a different pattern. Because the adversary selects future populations from the perturbed target-source library, methods that borrow information from the sources can sometimes become better aligned with the selected future direction as $r$ increases. Thus, GUARD, source--target mixing, and transGLM can show decreasing regret, but their absolute regret levels remained very different. As $r$ increased, GUARD's regret decreased from $0.009$ to $0.004$ for the centroid-centered adversary and from $0.018$ to $0.010$ for the target-centered adversary. In contrast, maximin's regret increased from $0.040$ to $0.068$ and from $0.042$ to $0.059$, respectively. Source--target mixing followed the same direction as GUARD under the centroid-centered adversary, decreasing from $0.030$ to $0.025$, but remained less accurate; its regret was $3.2$ times that of GUARD at $r=0.2$ and $6.0$ times that of GUARD at $r=1$. Pooled and RIFL had substantially larger regrets, approximately $1.0-1.5$. Overall, Design~I shows that GUARD preserves accuracy under several forms of post-target temporal drift, not only improving estimation at the current target.

\begin{figure}[H]
\centering
\includegraphics[width=1\textwidth]{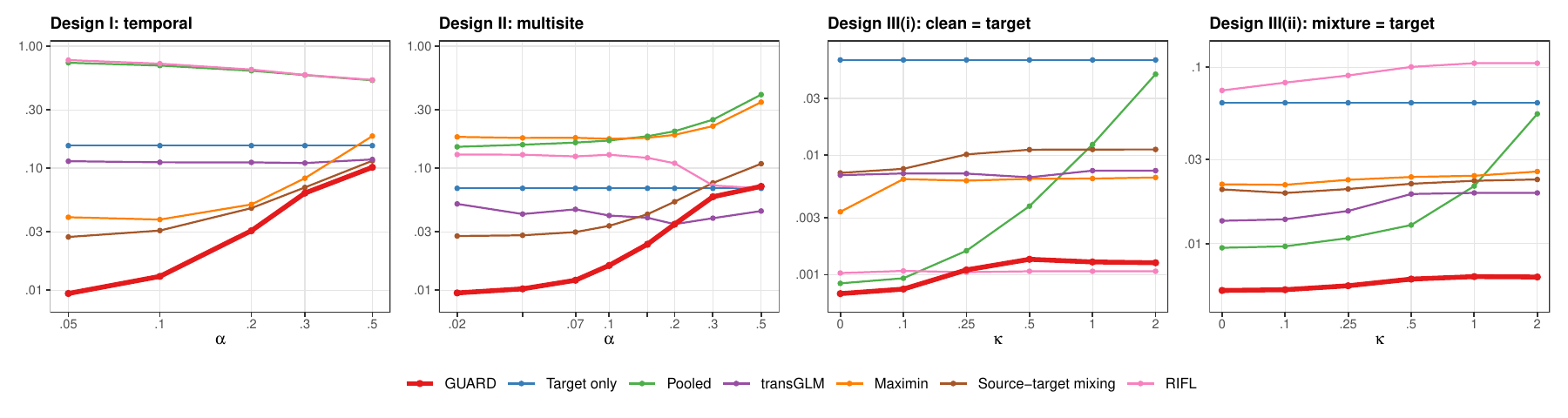}
\caption{\textbf{Estimation error} ($\Snorm{\hat\bbeta-\bbeta_0}$) under Design~I temporal domains and Design~II multisite transfer, as the target departs from the source convex hull ($\alpha$), and under Design~III sub-designs~(i) and~(ii), as the corrupted source moves farther from the target 
($\kappa$).}
\label{fig:studyB}
\end{figure}

We next evaluate current-target transfer learning under Design I-III. Although Design~I is used above to assess future generalization, it also provides a temporal-domain setting for evaluating current target estimation as source-mixing becomes less accurate.
Figure~\ref{fig:studyB} evaluates current-target estimation as $\alpha$ increases in Designs~I and~II, moving the target farther from exact source mixing, or as one source becomes increasingly corrupted in Design~III. Under exact source mixing ($\alpha=0$), GUARD attained an MSE of $0.0084$ in Design ~I, compared with $0.15$ for target only, and $0.0099$ in Design~II, compared with $0.028$ for source--target mixing and $0.068$ for target only. As $\alpha$ increased, GUARD's MSE rose gradually toward the target-only benchmark, but remained more accurate than other benchmarks under Design~I and the best transfer method across Design~II. In Design~II, source--target mixing ceased to improve over target only near $\alpha=0.3$, whereas GUARD approached target-only performance only near $\alpha=0.5$. The remaining methods did not exploit the mixture relationship to the same extent. Pooling was among the least accurate methods in both Designs~I and~II, because it ignored  heterogeneity in both the covariate and outcome spaces. RIFL also performed poorly in Design~I, consistent with its source-screening strategy not  tailored to settings where transferability holds primarily through a mixture of heterogeneous sources. TransGLM was less sensitive to changes in $\alpha$, but its MSE remained above that of GUARD throughout Design I, and when $\alpha\leq 0.2$ in Design II. Maximin did not adapt well to the target-specific deviation, especially in Design II, where MSE increased from $0.18$ to $0.35$. Thus, GUARD performed best when source mixing was exact or moderately violated and remained comparable to the best-performing methods even at larger $\alpha$.
Design~III assesses performance when sources except one corrupted (i) individually match the target; or (ii) can recover the target through a mixture. GUARD's MSE changed only from $0.0007$ to $0.0013$ as $\kappa$ increased from $0$ to $2$ in (i), and from $0.0054$ to $0.0065$ in (ii). Thus, GUARD remained stable and comparable to the best-performing methods even at larger $\kappa$. Pooling performed poorly in (i) because it failed to exclude the corrupted source. RIFL remained competitive in (i), but poorly in (ii). These results show that (i) favors source screening, whereas (ii) require mixture-based transfer; GUARD was best or close to the best in both settings.

\subsubsection{Inference Validity and Interval Quality}
\label{sec:studyC}

\begin{figure}[H]
\centering
\includegraphics[width=1\textwidth]{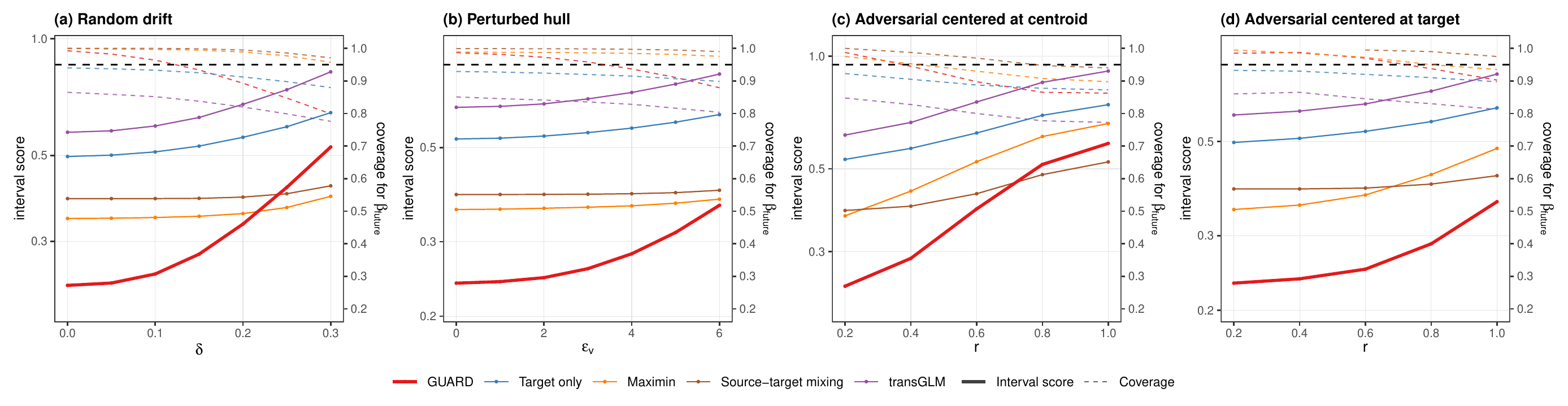}
\caption{\textbf{Interval quality under future drift.} IS  on each coordinate of $\bbetafut$ (solid, left axis) and coverage of $\bbetafut$ (dashed, right axis), under four different future distribution generators. Dashed line at the $0.95$ indicates nominal coverage level. Pooled and RIFL are omitted due to large IS ($3.1$--$4.3$).}
\label{fig:studyC}
\end{figure}

We next evaluate the performance of the re-sampling based uncertainty quantification. We first examine the performance of GUARD confidence intervals for $\bbeta\subGUARD(\lambda)$ at fixed values of the penalty parameter $\lambda$. Across the considered values of $\lambda$, GUARD maintained near-nominal coverage while producing substantially shorter intervals than target only. Detailed results are given in Appendix. We then assess interval quality for future generalization under Design~I using the interval score, followed by current-target inference under Designs~I--III.

Figure~\ref{fig:studyC} evaluates interval quality for future generalization under Design~I. Moving from left to right, the future target moves farther away from the current target. GUARD achieved the lowest interval score in most settings, indicating the best balance between interval length and coverage. At the current target, GUARD had the shortest average interval, with width 0.226, while maintaining empirical coverage 0.993. Its advantage was most pronounced under small-to-moderate random, perturbed-hull, and adversarial-hull drift. When the future target moved very far from the current target, GUARD’s shorter intervals became less conservative, and wider methods such as maximin or source--target mixing became competitive in some settings. Overall, GUARD provided the most favorable interval quality over the practically relevant range of future drift.

\begin{figure}[H]
\centering
\includegraphics[width=1\textwidth]{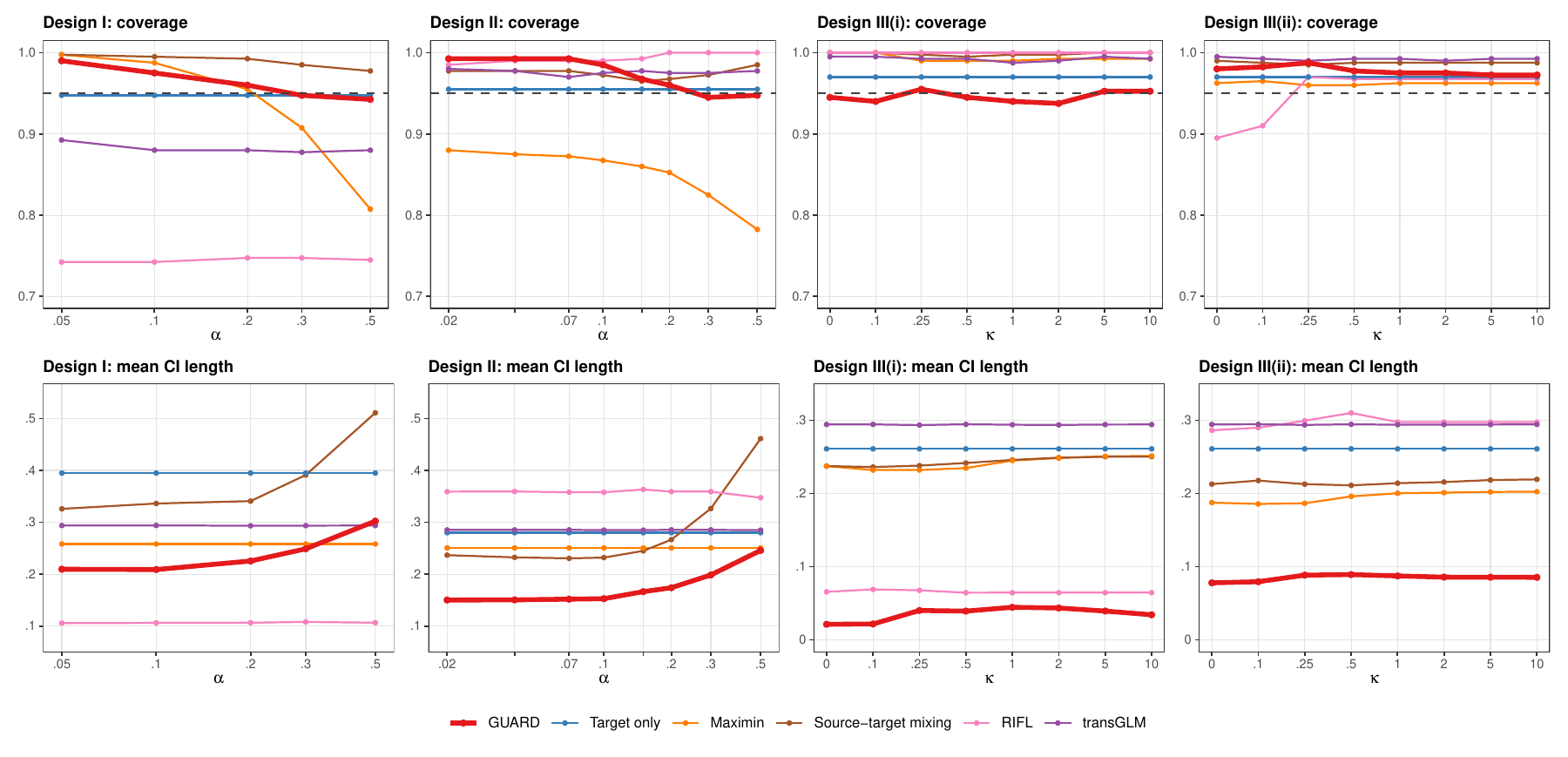}
\caption{\textbf{Current target inference for $\bbeta_0$.} Coverage of $\bbeta_0$ (top) and mean interval length (bottom), under different source-target generation mechanisms. }
\label{fig:studyExp}
\end{figure}

 Figure~\ref{fig:studyExp} evaluates the current target inference for$\beta_0$ under Designs I--III. In Designs I-II, the empirical coverage of GUARD remained close to the nominal level $95\%$ as $\alpha$ increased, while producing shortest intervals among methods that do produce valid coverage. For example, in Design I when $\alpha=0$, GUARD mean interval length was $0.21$, compared with $0.33$ for source--target mixing, $0.26$ for maximin, and $0.40$ for target only. On the other hand, RIFL and transGLM failed to attain valid coverage in Design~I, while in Design~II both attained the nomial coverage with the price of two largest intervals (transGLM 0.285; RIFL 0.347--0.363, against GUARD's 0.150--0.246).
In Design~III(i), where each clean source coincided with the target, GUARD, RIFL, and source--target mixing all maintained coverage. GUARD produced interval lengths of $0.021$--$0.045$, substantially shorter than target only (0.261), transGLM (0.293--0.295), source--target mixing (0.236--0.249), and RIFL (0.066--0.069). Overall, GUARD provided the most favorable coverage--length trade-off across the considered source--target structures.

\def\DAS{\mbox{DAS}_{3.2}}
\section{Robust Disease Activity Prediction in Rheumatoid Arthritis}\label{sec:real-design0}

Rheumatoid arthritis (RA) is a chronic inflammatory disease in which disease activity reflects the burden of joint and systemic inflammation. Because active inflammation can fluctuate over time and contribute to symptoms, functional limitation, and joint damage, RA care centers on longitudinal monitoring of disease activity. Modern treat-to-target management also uses repeated disease-activity assessment to guide treatment adjustment \citep{smolen2016treat_target,fraenkel2021acr_ra}. We study prediction of future CRP-based 28-joint Disease Activity Score (DAS), a composite measure of RA disease activity, using longitudinal EHRs. Accurate prediction of future DAS can help identify patients at risk of uncontrolled disease and support earlier clinical monitoring or treatment adjustment.

Deriving a temporally robust DAS prediction model using EHR data spanning multiple decades is challenging because patient population and clinical practice have shifted drastically over the years. Coding systems, laboratory ordering, medication availability, and treatment strategies all evolved over time. Diagnosis-code-derived features may shift around the ICD-9 to ICD-10 transition occurred around 2016, while treatment indicators may partly reflect the adoption of biologic and targeted synthetic disease-modifying antirheumatic drugs rather than only patient-level disease status. A model that pools historical years without accounting for these changes may therefore learn calendar-era patterns rather than stable predictors of future disease activity.

\subsection{Study Cohort, pre-trained feature representation, and evaluation design}
\label{sec:real-design}

Patient histories were aggregated into six-month windows using counts of EHR-derived features. For patient $i$ and window $t$, the prediction target was future DAS, $Y_{i,t}=\mathrm{DAS}_{i,t+1}$, modeled on the log scale; windows with observed future DAS were labeled and the rest unlabeled. Before applying GUARD, we pretrained a knowledge-graph-guided transformer on RA EHR data through 2009, comprising $33{,}329$ patients, $310{,}005$ patient-windows, and $3{,}816$ labeled windows. The model maps longitudinal EHR histories to patient-window embeddings, which were reduced by PCA to 22 dimensions while retaining about $95\%$ of the variation; the same transformation was applied to post-2009 windows. This representation captures relationships among related medications, laboratory tests, and other EHR concepts whose recording patterns may shift over time, providing a pretrained representation that GUARD can recalibrate using limited target labels and historical source years.

The downstream analysis included $671{,}003$ post-2009 windows from $62{,}719$ patients. The working design $\bA$ had $p_A=33$ columns, including demographics, CRP, ESR, the 22-dimensional embeddings, and the pretrained prediction $f_0(X)\); we also evaluated DML recalibration of $f_0(X)$ alone. The nuisance design $\bX$ additionally included four LLM-derived disease-severity indicators \citep{yang2026share_disease_activity}, used only in the outcome and density-ratio models.
We used calendar years to define target and source populations. For target year $t$, that year's windows formed the target and each preceding year formed a separate source. We evaluated $t=2012,\ldots,2017$, with evaluation through 2020. Across 2009--2020, there were $5{,}093$ labeled and $107{,}702$ unlabeled windows, reflecting a label-scarce setting. For each target year, $n_0=100$ labeled observations were used for training and the remainder for evaluation. To reduce labeled--unlabeled covariate shift, each labeled window was matched without replacement to 20 unlabeled windows using nearest-neighbor propensity-score matching, and only matched unlabeled windows were retained.

To evaluate prediction performance, we considered held-out samples from the current target year $(h\equiv v-t=0)$ and labeled windows from future years $(h>0)$, with the latter assessing robustness to temporal drift. For method $m$, target year $t$, and evaluation year $v\geq t$, we used the residual-variance-adjusted excess risk
\[
E_m^{t\to v}
=
{n_v}^{-1}\sum_{i\in\mathcal I_v}
\big(Y_i-\bA_i^{\top}\widehat{\bbeta}_{m}^{(t)}\big)^2
-
{(n_v-p_A)}^{-1}\sum_{i\in\mathcal I_v}
\big(Y_i-\bA_i^{\top}\widetilde{\bbeta}^{(v)}\big)^2,
\]
where $\widehat{\bbeta}_{m}^{(t)}$ is trained using target-year $t$ data and $\widetilde{\bbeta}^{(v)}$ is the least-squares fit in the evaluation year. Thus, $E_m^{t\to v}$ measures excess prediction error relative to an evaluation-year-specific fit. We also report Pearson correlation between predicted and observed DAS and AUC for classifying low disease activity, defined as DAS $<3.2$ \citep{fransen2009das_eular}. Metrics are averaged over all target-year/evaluation-year pairs with the same horizon $h$.

For uncertainty quantification, we evaluated coefficient-level CIs, mean-response intervals, and prediction intervals. Mean-response intervals target $\bA_i^\top\bbeta_0$ and were summarized across deciles of covariate extremity, while prediction intervals target future patient-level outcomes and were evaluated by empirical coverage and average length among records with observed outcomes. For GUARD, maximin, and source--target mixing, prediction intervals were constructed by adding an estimated residual variance to the resampling-based mean-response intervals before taking the union across stable resamples. Details on covariate-extremity stratification, residual variance estimation, and interval construction are provided in the Supplement.

\subsection{Results}
\subsubsection{Prediction performance}
\begin{figure}[h]
\centering
\includegraphics[width=1\linewidth]{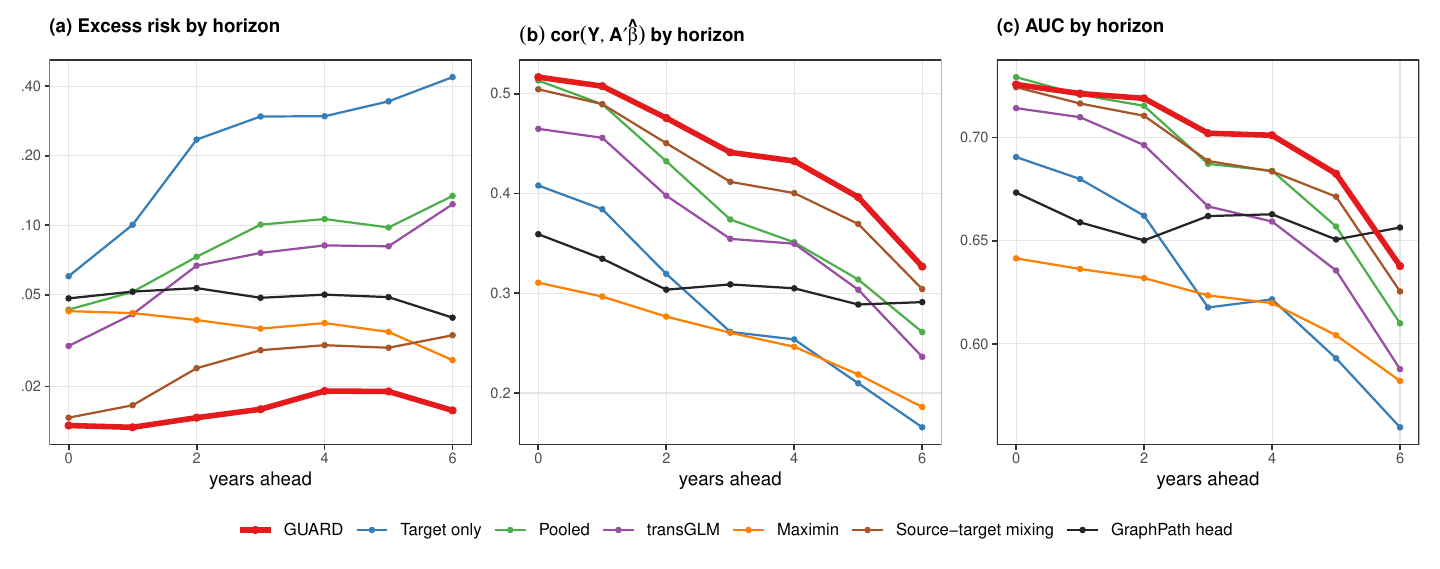}
\caption{Accuracy as a function of horizon, pooled over the six target years. Horizon~0 is the held-out target year. (a)~Excess risk on a log scale. (b) Pearson's correlation between DAS and predicted scores (c)~AUC for DAS below $3.2$.}
\label{fig:horizon}
\end{figure}

Figure~\ref{fig:target2012} summarizes performance with training target year 2012 across both current and future target years while Figure \ref{fig:horizon} summarizes average prediction performance across different horizons. As shown in Figure~\ref{fig:target2012}(a), with source data up to 2011 and training target year 2012, the model updated by GUARD attained the most robust performance. It exhibited little deterioration over this period: its excess risk was $0.0134$ on the held-out 2012 records and $0.0146$ eight years later. In contrast, the excess risk of target only estimator increased from $0.0699$ to $0.5232$, a $7.5$-fold increase; the corresponding increases were from $0.0216$ to $0.1914$ for pooling and from $0.0357$ to $0.1678$ for transGLM. Source-target mixing achieved a slightly lower excess risk than GUARD in the target year itself ($0.0114$ versus $0.0134$), but this advantage had disappeared by 2014, and its excess risk reached $0.0341$ in 2020. The recalibrated original $f_0(X)$ generally performed worse than GUARD or source mixing, suggesting that updating is necessary to improve performance. 

The horizon-level results in Figure~\ref{fig:horizon} show that GUARD's advantage persisted across training target years and horizons. Averaged over target-year and evaluation-year pairs with horizon $h$, GUARD's excess risk increased only from $0.014$ at $h=0$ to $0.016$ at $h=6$, a $16\%$ change. By comparison, excess risk increased by $130\%$ for source--target mixing, $210\%$ for pooling, $310\%$ for transGLM, and $630\%$ for target only. Although Maximin and the recalibrated model $f_0(X)$ were relatively stable across horizons, their absolute excess risk remained higher than GUARD's; at $h=6$ their excess risks were $0.026$ and $0.040$, compared to $0.016$. GUARD also retained substantially higher $R^2$ and AUC across horizons, except for $h=6$ where the original GraphPath head achieved the best AUC. Overall, GUARD achieved both low excess risk and strong resistance to temporal drift, more effectively translating the pretrained representation into accurate DAS prediction over time.

\subsubsection{Uncertainty Quantification}
\begin{figure}[htbp]
\centering
\includegraphics[width=1\linewidth]{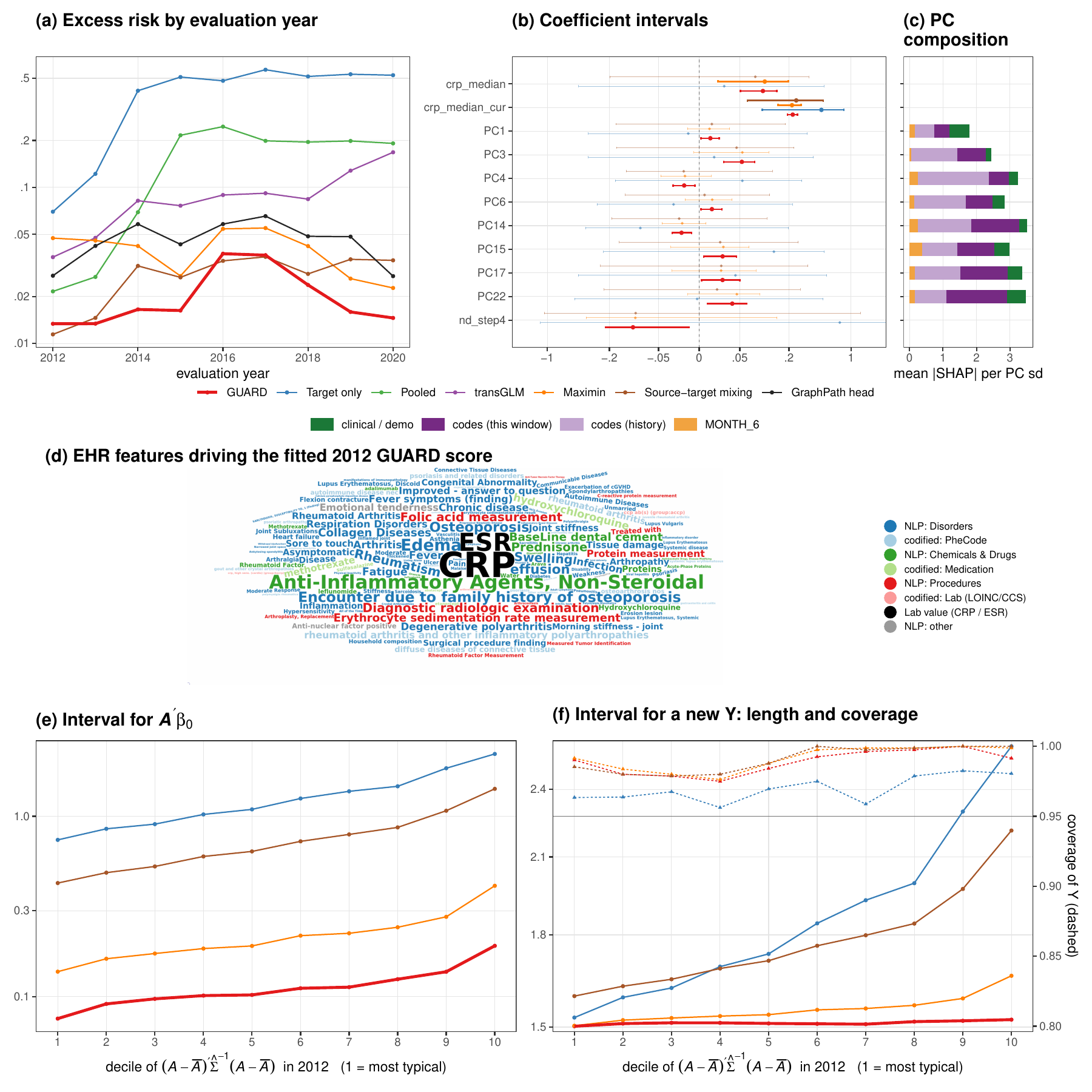}
\caption{Target-year 2012 analysis. Results are averaged over $20$ random splits. (a) Excess risk across evaluation years. (b) Coefficient confidence intervals. (c) Embedding direction feature-block composition. (d) Wordcloud of raw EHR features contributing to GUARD’s 2012 DAS predictions, with font size proportional to the SHAP value. (e) Mean-response interval length and (f) prediction-interval length and coverage across covariate-extremity deciles.}
\label{fig:target2012}
\end{figure}

GUARD also produced sharper coefficient-level inference as shown in Figure~\ref{fig:target2012}b. For target year 2012, the average confidence-interval length across the $33$ coordinates was $0.101$ for GUARD, compared with $0.234$ for maximin, $1.304$ for source--target mixing, and $2.306$ for target only. Thus, intervals from source--target mixing and target only were $12.9$ and $22.8$ times longer, respectively, than those from GUARD. This precision allowed to GUARD to detect $11$ coordinates, including 8 embeddings, with coefficients significantly different from 0, compared with 2 for maximin and 1 each for source--target mixing and target only, both only identifying current-window median CRP. Figure~\ref{fig:target2012}(c) summarizes the EHR feature blocks underlying the eight embedding directions selected by GUARD, using SHAP values normalized by the standard deviation of each component \citep{lundberg2017shap}. These selected directions were driven mainly by longitudinal EHR code patterns, which accounted for $80.4\%$ of the total attribution, including $44.6\%$ from historical counts and $35.8\%$ from the current window. Demographic and clinical variables contributed $13.0\%$. Figure~\ref{fig:target2012}(d) summarizes key contributors to the fitted 2012 GUARD score as measured by their normalized SHAP vales. Two key laboratory measurements for RA activity, CRP and ESR, accounted for $45.7\%$ of the displayed attribution. Narrative concepts accounted for $48.7\%$, while codified EHR variables accounted for the remaining $5.6\%$. The most highly ranked concepts, including nonsteroidal anti-inflammatory agents, edema, joint effusion, swelling, prednisone, and rheumatism, are clinically consistent with active RA and its treatment. Thus, GUARD effectively leverages the semantic structure learned by the pretrained representation while emphasizing stable, recognizable RA signals such as inflammatory markers, symptoms, and treatments.

Mean-response intervals also clearly showed GUARD's uncertainty gain (Figure~\ref{fig:target2012}e). From covariate-extremity decile~1 to decile~10, GUARD's interval length increased from $0.076$ to $0.191$, compared with $0.740$ to $2.216$ for target only. GUARD's intervals were therefore $9.8$ and $11.6$ times shorter at the most typical and most extreme profiles, respectively. They also widened less with covariate extremity, increasing by a factor of $2.53$, compared with $2.99$ for both maximin and target only and $3.33$ for source--target mixing.

Prediction intervals show a different pattern from the mean-response intervals because they include both estimation uncertainty and residual outcome variation. As shown in Figure~\ref{fig:target2012}(f), all methods achieved approximately nominal coverage across covariate-extremity deciles, but their interval lengths diverged for more extreme patient profiles. GUARD’s prediction intervals remained nearly flat across deciles, increasing only from about 1.50 for typical profiles to 1.52 for the most extreme profiles. In contrast, target only widened substantially, reaching 2.61 at decile~10, and source--target mixing also became noticeably wider. Thus, even when residual variation dominates prediction uncertainty, GUARD maintains valid coverage while avoiding the large interval inflation seen for less stable target-only or transfer estimators.

\section{Discussion}\label{sec:discussion}

We have introduced GUARD, a semi-supervised framework for post-deployment model updating that anchors a robust update to the current target population while calibrating source borrowing to estimation uncertainty. Anchoring in this way keeps the target labels informative without giving up protection against future drift, and the uncertainty matrix $V$ directs borrowing toward sources that can be estimated well. The simulations and the RA EHR application show that this translates into accuracy that holds up as the deployment environment moves away from the training target, with more reliable intervals than competing methods.

We assume throughout that labels are missing completely at random within each  population. In EHR, this is reasonable by design when labels are curated via random sampling of patients for chart review. The GUARD framework can accommodate labeling that is missing at random given $X$ by introducing a selection propensity $\pi_\ell(x) =
\Pr(R = 1 \mid X = x, \text{population } \ell)$ and weighting the labeled contributions in \eqref{eq:source_beta_hat_main} by $\pi_\ell(x)^{-1}$, at the cost of one additional nuisance function per population and a correspondingly modified $\theta_\ell$ in \eqref{eq:theta_exact}. In the application we mitigate labeled--unlabeled covariate imbalance by propensity matching each labeled window to twenty unlabeled windows, but a formal treatment of outcome-dependent labeling warrants future research.

The population quantities, the coefficient library $B$, the uncertainty matrix $V$, and the transferable set in Definition~\ref{def:transferability}, depends
on the feature map $A=A(X)$. Thus, transferability depends on the
representation as much as on the populations. This makes the choice of $A$ an important design decision. It must be rich enough for a linear update to correct the drift, but small enough to fit from a few hundred target labels. In our EHR application, we use a pretrained representation via a transformer architecture followed by dimension reduction via principal components, plus the deployed score $f_0(X)$ and a few clinical covariates. A pretrained representation with a projection head is a reasonable starting point when a deployed model already exists, but building a theoretically justified representation step into GUARD itself warrants future research.

We work throughout in the regime $N_0 \gg \max_{0 \le \ell \le L} n_\ell$, under which the shared target-side term $R$ in \eqref{eq:cor1-V-matrix} is of smaller order than the diagonal labeled-data term $D$, so that $V$ is approximately diagonal. When this regime is unavailable, one may instead partition the target sample into $L$ disjoint subsets and use the $\ell$th subset for all target-side quantities entering $\widehat\beta_\ell$. If each source also uses
its own independent data, the columns of $\widehat B$ are then built from mutually independent blocks and $V$ is diagonal by construction. The cost is that each block carries roughly $n_0 / L$ target observations, so $\Sigma_0$ and the target column are estimated less precisely; the regularity conditions of Section~\ref{sec:theory} must be restated in terms of the per-block sample size.

Although our theoretical analysis targets the robust prediction model
$\beta_{\mathrm{GUARD}}^\star(\lambda)=B\gamma_\lambda^\star$,
the current algorithm can also support inference for the target
coefficient $\beta_0$ under additional conditions. For a prespecified
contrast $a$, the same screened union confidence interval remains
asymptotically valid for $a^\top\beta_0$ if
$|a^\top\{\beta_{\mathrm{GUARD}}^\star(\lambda)-\beta_0\}|
=o\{\sigma_a^\star(\lambda)\}$, where
$\sigma_a^\star(\lambda)$ denotes the standard deviation of the
leading fixed-weight estimation error
$a^\top(\widehat B-B)\gamma_\lambda^\star$, and the conditions
ensuring coverage of the GUARD target hold.
When inference concerns $\beta_0$, $\lambda$ therefore serves as
a tuning parameter for controlling transfer bias relative to the
inferential scale, where a sufficient large $\lambda$ may ensure $|a^\top\{\beta_{\mathrm{GUARD}}^\star(\lambda)-\beta_0\}|
=o\{\sigma_a^\star(\lambda)\}$. 

The GUARD construction rests on the identity $\beta_{\mathrm{new}}(\gamma) = B\gamma$. Under squared loss, a convex mixture of conditional outcome laws induces the \emph{same} convex mixture of target-projected coefficients, which is what reduces (\ref{eq:guard_dro_primary}) to a quadratic program on the simplex and makes $V$ the natural object to penalize. This fails under cross-entropy loss, where the optimal linear predictor under a mixture is not the mixture of the population-specific optimal predictors. Thus extending to non-linear link would need a local quadratic expansion around $\beta_0$ or a direct treatment of the conditional
group DRO problem in the spirit of \citet{guo2025statistical}. In this paper, we consider mixture distributions $\Pnew(\gamma)$ for the conditional laws while holding $\Pt_X$ fixed, so GUARD is robust to concept drift but not to future covariate shift. This keeps the estimand on a single interpretable scale, since $\beta_0$ is an $L_2$ projection under $\Pt_X$ and would itself move otherwise. It may be of interest to further incorporate bounded perturbations of $\Pt_X$, similar to \citet{kim2026distributionally}, which could further widens the uncertainty set at the cost of a moving target of inference. 

The framework extends to a high-dimensional working feature map under sparsity and regularity conditions, with two modifications. First, when the target-projected coefficients $\beta_0, \dots, \beta_L$ are sparse, the unregularized projection estimator in \eqref{eq:source_beta_hat_main} is replaced by
\begin{equation}
\widehat\beta_\ell \in \arg\min_{b \in \mathbb{R}^p}
\Bigl\{ b^{\mathsf{T}} \widehat\Sigma_0 b - 2 b^{\mathsf{T}} \widehat\mu_\ell
+ \rho_\ell \|b\|_1 \Bigr\}, \qquad \ell = 1, \dots, L,
\label{eq:sparseproj}
\end{equation}
where $\rho_\ell > 0$ is a regularization parameter distinct from the GUARD penalty $\lambda$, and $\widehat\beta_0$ is estimated by the Lasso on the labeled target data. This requires a restricted eigenvalue condition on $\Sigma_0$ and continues to treat $L$ as fixed. Second, the inferential target is no longer the full vector but a subvector $\beta_{0,S}$ with $S \subset \{1, \dots, p\}$, or a low-dimensional transformation of it; the corresponding uncertainty matrix has entries $V_{\ell\ell'} = \mathbb{E}[(\widehat\beta_{\ell,S} - \beta_{\ell,S})^{\mathsf{T}}\Sigma_{0,SS} (\widehat\beta_{\ell',S} - \beta_{\ell',S})]$, computed from standard debiased Lasso results.

GUARD fits the regulatory setting that motivated this work: it holds the approved rule $f_0$ fixed, updates only a low-dimensional head in a
prespecified feature space from a few newly labeled observations, and reports intervals for the updated quantity. Because the update depends only on prespecified inputs---$A$, the source definitions, and $\lambda$---it can be written into a predetermined change control plan. One gap remains. A health system would update recursively, with this year's model becoming next year's $f_0$ and this year's target becoming next year's source, which breaks the cross-population independence in Assumption~\ref{cond:setup}. Characterizing error accumulation across repeated cycles is an open problem, and an important one for keeping clinical algorithms reliable over decades.

\bibliography{ref}
\end{document}